\documentclass[twocolumn,showpacs,amsmath,amssymb,superscriptaddress]{revtex4}

\usepackage{dcolumn}

\usepackage{graphicx}
\usepackage{dcolumn}
\usepackage{bm}

\begin{document}


\title{Collective structural evolution in neutron-rich Yb, Hf, W, Os and
Pt isotopes}

\author{K.~Nomura}
\affiliation{Department of Physics, University of Tokyo, Hongo,
Bunkyo-ku, Tokyo 113-0033, Japan} 

\author{T.~Otsuka}
\affiliation{Department of Physics, University of Tokyo, Hongo,
Bunkyo-ku, Tokyo 113-0033, Japan} 
\affiliation{Center for Nuclear Study, University of Tokyo, Hongo,
Bunkyo-ku, Tokyo 113-0033, Japan} 
\affiliation{National Superconducting Cyclotron Laboratory, 
Michigan State University, East Lansing, Michigan 48824-1321, USA}

\author{R.~Rodr\'\i guez-Guzm\'an}
\affiliation{Instituto de Estructura de la Materia, IEM-CSIC, Serrano
123, E-28006 Madrid, Spain} 

\author{L.~M.~Robledo}
\affiliation{Departamento de F\'\i sica Te\'orica, Universidad
Aut\'onoma de Madrid, E-28049 Madrid, Spain}

\author{P.~Sarriguren}
\affiliation{Instituto de Estructura de la Materia, IEM-CSIC, Serrano
123, E-28006 Madrid, Spain} 
\date{\today}

\begin{abstract}

An interacting boson model Hamiltonian determined from 
Hartree-Fock-Bogoliubov calculations with the new microscopic Gogny 
energy density functional D1M, is applied to the spectroscopic 
analysis of neutron-rich Yb, Hf, W, Os and Pt isotopes with mass 
$A\sim 180-200$. Excitation energies and transition rates for the 
relevant low-lying quadrupole collective states are calculated by 
this method. Transitions from prolate to oblate ground-state shapes 
are analyzed as a function of neutron number $N$ in a given isotopic 
chain by calculating excitation energies, $B$(E2) ratios, and 
correlation energies in the ground state. It is shown that such 
transitions tend to occur more rapidly for the isotopes with lower 
proton number $Z$, when departing from the proton shell closure 
$Z=82$. The triaxial degrees of freedom turn out to play an 
important role in describing the considered mass region. Predicted 
low-lying spectra for the neutron-rich exotic Hf and Yb isotopes are 
presented. The approximations used in the model and the 
possibilities to refine its predictive power are addressed. 

\end{abstract}

\pacs{21.10.Re,21.60.Ev,21.60.Fw,21.60.Jz}

\maketitle



\section{Introduction\label{sec:introduction}}

The study of the origin of nuclear deformation 
and its evolution as a function of
proton and neutron numbers has attracted considerable 
theoretical interest from a large variety of viewpoints
\cite{BM,RS,review,Werner,Bender_review,Heenen-nature,Rayner-1,rayner-PRL,
Naza-def,Sarriguren2008Sk,Robledo-1,Mottelson,Cejnar_Jolie_Casten_review}.
Experimentally, low-lying spectroscopy provides a very powerful source
of information that allows one to establish signatures correlating the
nuclear shape evolution with the energy spectra \cite{Julin,draco,Davidson1994Os,
Davidson1999Pt,Kibedi1994Os,Kibedi2001W,wu96,podolyak,pfun,caamano}.


Among many other nuclear structure models, self-consistent 
mean-field methods, based on microscopic energy density functionals 
(EDFs), have provided both accurate and universal descriptions of 
different nuclear intrinsic properties including binding energies, 
ground-state deformations, density distributions, low-lying 
one-quasiparticle configurations, as well as the way nuclear shapes 
evolve with the number of nucleons 
\cite{RS,Bender_review,Robledo-1,Sarriguren2008Sk,Rayner-1,rayner-PRL, 
odd-rayner-2,odd-rayner-1,odd-rayner-3,schunck,pot10}. 
Popular EDFs are the non-relativistic Skyrme \cite{Bender_review,Sk,VB} 
and Gogny \cite{Go,gogny-other} ones, as well 
as relativistic mean-field 
Lagrangians \cite{RMF_review}. To describe nuclear spectroscopy one 
should go beyond the mean-field approximation to take into account 
the restorations of broken symmetries and/or the configuration 
mixing of intrinsic states in the spirit of the generator coordinate 
method (GCM) 
\cite{RS,Bender_review,Bender2006correlation,Bender-triaxial,NPA-rayner,
Rayner-Pb-GCM,Yao2011GCM2}. In this kind of studies calculations may 
become computationally much more demanding and time consuming than the underlying mean
field, particularly when triaxial degrees of freedom are included in the analysis. 


A sound approximation to the full GCM configuration mixing and/or the 
symmetry restoration is the five-dimensional collective 
Hamiltonian with quadrupole degrees of freedom where both rotational and 
vibrational mass parameters are determined from the constrained, 
self-consistent mean-field calculations with a given EDF and the 
collective potential is derived by the zero-point energy correction 
to the total mean-field energy (e.g., \cite{CollSk,CollGo,CollRHB}). 

Alternatively, nuclear dynamics and spectroscopic quantities can be 
approximated by introducing appropriate bosonic degrees of freedom. 
The interacting boson model (IBM) \cite{IBM} can be regarded as a 
nice example for this, and has been exploited in a large number of 
phenomenological studies focusing on the low-lying spectrum of 
medium-heavy and heavy nuclei \cite{IBM}. The simplest version of 
the IBM is built on monopole $s$ and quadrupole $d$ bosons, which 
reflect the collective $J^{\pi}=0^{+}$ and $2^{+}$ pairs of 
valence-shell configurations, respectively \cite{OAI}. Nevertheless, 
since the IBM itself should have a certain microscopic foundation, a 
Hamiltonian of IBM has been derived conventionally from the 
shell-model configuration \cite{OAI}, and more recently from 
EDF-based calculations \cite{nso}. These mapping methods have been 
applied to realistic cases involving a variety of situations
covering from nuclei with  modest 
quadrupole deformation including $\gamma$ -unstable ones 
\cite{OAI,MO,nso,nsofull,GognyIBMPt}, to strongly deformed rotational 
nuclei \cite{gboson-1,gboson-2,OPLB138,OY,IBMrot,IBMwos}. Also 
quantum-mechanical correlation effects in the ground state 
have been considered \cite{nsofull}. Starting from the constrained 
Hartree-Fock-Bogoliubov (HFB) theory with the D1S \cite{D1S} 
parametrization of the Gogny functional, the method of \cite{nso} 
was used for the spectroscopic analysis of Pt isotopes 
\cite{GognyIBMPt}, and some of Os and W isotopes \cite{IBMwos}. 


In this paper we apply the mapping procedure of \cite{nso} to the 
mass region $A\sim 180-200$, extending the analysis made in 
\cite{GognyIBMPt,IBMwos} to the neighboring exotic Hf and Yb nuclei. An 
additional motivation is to explore some possibilities to refine the 
predictive power of the method for the considered mass region. 
Although the D1S parametrization of the Gogny force is considered as 
a global EDF able to describe many low-energy nuclear data with 
reasonable predictive power (see, for example, 
Refs.~\cite{CollGo,RaynerPt,gradient-2} and references therein), we have 
preferred to use in this paper the Gogny-D1M functional \cite{D1M}. 
Systematic explorations of different nuclear phenomena \cite{RaynerPt,D1M}, 
including properties of odd nuclei computed within 
the equal-filling approximation \cite{odd-rayner-1,odd-rayner-2,odd-rayner-3}, 
suggest that the new 
parametrization of the Gogny-EDF is as good as the standard D1S, a 
fact that we intent to confirm in this paper.


It is well known that the nuclei in the mass region $A\sim 180-200$ 
exhibit a transition between prolate and oblate equilibrium shapes 
as a function of the nucleon number, with the critical point around 
$N\approx 116$ having a pronounced $\gamma$ softness 
\cite{Jolie2003HgHf,Pt196O6,CastenCizewski,198Os}. These facts make the 
region a potential testing ground  to understand the deformation 
properties of atomic nuclei. The evolution of the nuclear ground 
states in this mass region has been investigated recently with the 
constrained self-consistent mean-field method with microscopic EDFs 
\cite{Sarriguren2008Sk,gradient-2,RMFPt}. Both the (constrained) 
Hartree-Fock+BCS (HF+BCS) and the HFB approximations have been used 
to compute energy surfaces with quadrupole degrees of freedom in 
order to give a microscopic insight into shape transitions 
\cite{gradient-2,RaynerPt,Sarriguren2008Sk}. It was shown in these 
studies that the triaxiality is an important ingredient to describe 
the evolution from prolate to oblate shapes, irrespective of the 
types of the EDFs used. 

It should be kept in mind that Pt, Hg, and Pb isotopes are well 
known \cite{Andre} for the spectacular coexistence of different 
low-lying configurations based on different intrinsic deformations 
as observed in their low lying spectrum. There are a number of works 
aimed at the understanding of the shape coexistence phenomenon in 
this region in terms of both EDF-based microscopic calculations 
\cite{Rayner-Pb-GCM,DuguetGCMPb,BenderGCMPb} and phenomenological models 
\cite{CMIBM,IBM1-Pt-MCZ,IBMCMECQF}.

The paper is organized as follows: In Sec.~\ref{sec:theory}, a short 
outline of the theoretical framework is given. 
Section~\ref{sec:results} presents the energy surfaces, ground-state correlation 
energies, moments of inertia for the rotational bands, low-lying 
spectra, and the $B$(E2) systematics for the considered isotopes 
chains. Section~\ref{sec:summary} is devoted to the concluding 
remarks and work perspectives.

\section{Theoretical procedure}
\label{sec:theory}

The analysis starts with a constrained HFB calculation using the 
Gogny-D1M EDF. The constraints in this case refer to the mass 
quadrupole moments which are associated with the quadrupole 
deformation parameters $\beta$ and $\gamma$ in the geometrical model 
\cite{BM}. The set of constrained HFB calculations, for 
each collective coordinate ($\beta$,$\gamma$), provides  the total HFB 
energy (denoted by $E_{\rm HFB}(\beta,\gamma)$). For calculation 
details the reader is referred to \cite{gradient-2,GognyIBMPt}. 

In other studies solving the five-dimensional collective Hamiltonian 
\cite{CollSk,CollGo,CollRHB}, the collective potential energy surface is obtained 
by subtracting the zero-point energies for both rotational and 
vibrational motions from the constrained HFB energy surface. This 
corrected energy surface should be viewed as a collective potential 
energy surface. In the present work, the constrained HFB energy
surface and the corresponding boson energy surface are
compared, and they will be referred to simply as {\em energy
surface}. Note 
that, as the total energy is considered, all ingredients including those 
relevant to kinetic terms are supposed to be taken into account to a 
good extent.  

Each point of  the Gogny-HFB energy surface $E_{\rm 
HFB}(\beta,\gamma)$ is mapped onto the corresponding point on the 
bosonic energy surface, denoted by $E_{\rm 
IBM}(\beta_{B},\gamma_{B})$ with $\beta_{B}$ and $\gamma_{B}$ being 
the deformation parameters for the boson system, in such a way that 
the bosonic energy surface fits the fermionic one \cite{nso}. In 
this paper we consider the proton-neutron interacting boson model 
(IBM-2) \cite{OAI} because it reflects better the microscopic 
picture than the original version of the IBM without distinction of 
the proton and the neutron degrees of freedom (often called IBM-1). 
In what follows we denote the IBM-2 simply as the IBM, unless 
otherwise specified. The IBM energy surface is obtained as the 
expectation value of a given boson Hamiltonian \cite{coherent} in 
terms of the coherent state $|\Phi(\beta_{B},\gamma_{B})\rangle$. 
The coherent state represents the intrinsic wave function of the 
boson system, and is characterized by the deformation variables 
$\beta_{B}$ and $\gamma_{B}$. In principle, proton and neutron 
bosons might have different values of the deformation parameters, 
but since proton and neutron systems are supposed to attract each 
other strongly in medium-heavy and heavy deformed nuclei, the 
deformations of proton and neutron systems can be taken the same to a 
good approximation. 


If the separability of the mapping along the $\beta$ and the $\gamma$
directions is assumed, one can consider the relation between the 
IBM and the geometrical deformation variables \cite{nso,nsofull}. It 
was shown \cite{coherent} that, in general terms, the bosonic and the 
geometrical $\beta$s are proportional to each other and that the 
proportionality coefficient coincides with the ratio of the 
total nucleon number to the valence nucleon number counted from the 
nearest closed shells. We exploit this relation and assume that 
$\beta_{\rm B}=C_{\beta}\beta$, with $C_{\beta}$ being a numerical 
coefficient \cite{nso}. The typical range of the $C_{\beta}$ value 
turns out to be approximately $5\sim 10$, which is about the same 
order of magnitude as the actual ratios of the total nucleon number 
to the valence nucleon number. Regarding the triaxial parameter $\gamma$, 
the identification $\gamma_{\rm B}=\gamma$ seems  valid as indeed both 
geometrical and IBM $\gamma$'s have the same meaning, ranging from 0 
to 60 degrees. 

We adopt the IBM Hamiltonian of the following form: 
\begin{eqnarray}
\hat H_{\rm IBM} = \epsilon \hat n_{d}+\kappa
\hat Q_{\pi}\cdot \hat Q_{\nu} + \alpha\hat L\cdot\hat L,  
\label{eq:bh}
\end{eqnarray} 
where the first term $\hat n_{d} = \hat n_{d \pi}+\hat n_{d \nu}$ with 
$\hat n_{d \rho}=d_{\rho}^{\dagger}\cdot\tilde d_{\rho}$ 
($\rho=\pi$ or $\nu$) is identified as the $d$-boson number operator. 
The second term on the right-hand side of Eq.~(\ref{eq:bh}) stands for
the quadrupole-quadrupole interaction between 
proton and neutron systems, with 
$\hat Q_{\rho}=s_{\rho}^{\dagger}\tilde d_{\rho}+d_{\rho}^{\dagger}\tilde s_{\rho}+\chi_{\rho}[d_{\rho}^{\dagger}\tilde d_{\rho}]^{(2)}$ 
being the quadrupole operator for proton or neutron systems.
The third term (denoted by LL term, hereafter) is
relevant to the moment of inertia of 
the rotational band. $\hat L=\hat L_{\pi}+\hat L_{\nu}$ is the
angular momentum operator for the boson system with 
$\hat L_{\rho}=\sqrt{10}[d_{\rho}^{\dagger}\tilde d_{\rho}]^{(1)}$. 

The form of the Hamiltonian $\hat H_{\rm IBM}$ in Eq.~(\ref{eq:bh}) is not 
the most general,
but embodies all essential features of the low-lying quadrupole collective states. 
A more general IBM Hamiltonian with up to two-body interactions contains
many more terms  than those considered here. 
However, these additional terms are supposed to be of little
importance, and their implementation would increase the number of parameters,
which makes the problem quite complicated. 

\begin{table}[htb!]
\caption{\label{tab:IBMpara}%
The parameters for the IBM Hamiltonian $\hat H_{\rm IBM}$ of
 Eq.~(\ref{eq:bh}), as well as the coefficient $C_{\beta}$, obtained from the mapping of HFB to IBM energy
 surfaces for the considered Yb, Hf, W, Os and Pt nuclei with $N=110-122$. 
}
\begin{ruledtabular}
\begin{tabular}{ccccccc}
\textrm{}&
\textrm{$\epsilon$}&
\textrm{$-\kappa$}&
\textrm{$\chi_{\pi}$}&
\textrm{$\chi_{\nu}$}&
\textrm{$\alpha$}&
\textrm{$C_{\beta}$}\\
\textrm{}&
\textrm{(keV)}&
\textrm{(keV)}&
\textrm{$\times 10^{3}$}&
\textrm{$\times 10^{3}$}&
\textrm{(keV)}&
\textrm{}\\
\colrule
$^{180}$Yb & 212 & 265 & 337 & -991 & -9.06 & 3.60 \\
$^{182}$Yb & 169 & 265 & 300 & -900 & -11.4 & 3.70\\
$^{184}$Yb & 279 & 271 & 302 & -548 & -9.84 & 3.87\\
$^{186}$Yb & 418 & 268 & 147 & -106 & -9.54 & 4.90\\
$^{188}$Yb & 528 & 265 & 418 &   43 & -4.68 & 5.13\\
$^{190}$Yb & 769 & 267 & 332 &  573 & -0.185 & 5.50\\
$^{192}$Yb & 806 & 271 & 461 &  862 & 21.5 & 7.20\\
\colrule
$^{182}$Hf & 124 & 280 & 489 & -913 & -5.61 & 3.93 \\
$^{184}$Hf & 128 & 282 & 458 & -938 & -8.01 & 4.07\\
$^{186}$Hf & 109 & 275 & 400 & -700 & -4.85 & 4.40\\
$^{188}$Hf & 250 & 277 & 282 & -208 & -7.90 & 5.30\\
$^{190}$Hf & 442 & 280 & 403 &  -30 & -5.99 & 5.48\\
$^{192}$Hf & 619 & 273 & 388 &  443 & 2.79 & 5.94\\
$^{194}$Hf & 716 & 277 & 534 &  805 & 18.4 & 8.20\\
\colrule
$^{184}$W & 50.4 & 286 & 409 & -859 &  -0.400 & 4.09 \\
$^{186}$W & 36.8 & 285 & 389 & -835 &  -2.30 & 4.50\\
$^{188}$W & 69.6 & 289 & 401 & -662 &  -1.44 & 4.80\\
$^{190}$W & 71.3 & 275 & 572 & -419 &  -2.72 & 5.60\\
$^{192}$W & 231 & 270 & 189 & 147 &  -4.15 & 6.30\\
$^{194}$W & 627 & 291 & 392 & 536 &  -5.74 & 6.87\\
$^{196}$W & 686 & 281 & 745 & 822 &  15.3 & 8.50\\
\colrule
$^{186}$Os & 142 & 310 & 331 & -689 & -0.433 & 4.40 \\
$^{188}$Os & 162 & 318 & 352 & -672 & -2.78 & 4.83\\
$^{190}$Os & 86.7 & 303 & 412 & -509 & -2.61 & 5.40\\
$^{192}$Os & 91.5 & 292 & 502 & -488 & -3.09 & 6.15\\
$^{194}$Os & 289 & 305 & 401 & -77 & -6.04 & 6.74\\
$^{196}$Os & 541 & 298 & 336 & 513 & -5.94 & 7.64\\
$^{198}$Os & 683 & 304 & 573 & 793 & 8.50 & 9.66\\
\colrule
$^{188}$Pt & 187 & 328 & 409 & -487 & 8.16 & 4.81 \\
$^{190}$Pt & 215 & 336 & 300 & -10 & 5.93 & 5.56\\
$^{192}$Pt & 311 & 362 & 265 &  44 & -0.117 & 6.44\\
$^{194}$Pt & 312 & 366 & 490 & -50 & 0.214 & 6.85\\
$^{196}$Pt & 435 & 356 & 475 & 311 & 1.87 & 7.28\\
$^{198}$Pt & 489 & 319 & 611 & 565 & 8.80 & 7.90\\
$^{200}$Pt & 719 & 308 & 467 & 949 & -4.69 & 8.78\\
\end{tabular}
\end{ruledtabular}
\end{table}

The parameters contained in the first two terms of the Hamiltonian 
$\hat H_{\rm IBM}$ in Eq.~(\ref{eq:bh}), $\epsilon$, $\kappa$, 
$\chi_{\pi}$ and $\chi_{\nu}$, as well as the coefficient $C_{\beta}$, 
are fixed using the fitting method of Ref.~\cite{nsofull}. The LL 
term contributes to the energy surface in the same way as the $d$
-boson number operator, but with a different coefficient, $6\alpha$. 
Hence, the $\alpha$ coefficient cannot be fixed only by the mapping of the 
energy surface. A further step is required, in order to incorporate 
specific non-zero angular frequency features of the rotational cranking. 
The $\alpha$ value is determined by the procedure 
of Ref.~\cite{IBMrot}, where the cranking moment of inertia was 
compared between fermion and boson systems. 

We then calculate the moment of inertia for the $2^{+}_{1}$ excited
state by the Thouless and Valatin (TV) formula \cite{TV}, 
\begin{eqnarray}
 {\cal J}_{\rm TV}=3/E_{\gamma}. 
\label{eq:TV}
\end{eqnarray}
Here, $E_{\gamma}$ stands for the $2^{+}_{1}$ excitation energy obtained from 
the self-consistent cranking method with the constraint 
$\langle\hat J_{x}\rangle=\sqrt{L(L+1)}$, where $\hat J_{x}$ represents
the $x$-component of the (fermion) angular momentum operator \cite{gradient-2}. 
In \cite{IBMrot}, the Inglis-Belyaev formula
\cite{Inglis,Belyaev} turned out to be valid for the rotational regime,
but the present TV moment of inertia appears to be more general.  

For the boson system, we calculate the moment of inertia
of the intrinsic (coherent) state, denoted by ${\cal J}_{\rm IBM}$,  using the cranking
formula of Ref.~\cite{IBMcranking} 
\begin{eqnarray}
 {\cal J}_{\rm IBM}(\beta_{B},\gamma_{B})=\lim_{\omega\to 0}
\frac{1}{\omega}
\frac{\langle\Phi(\beta_{B},\gamma_{B})|\hat L_{x}|\Phi(\beta_{B},\gamma_{B})\rangle}{\langle\Phi(\beta_{B},\gamma_{B})|\Phi(\beta_{B},\gamma_{B})\rangle}, 
\label{eq:IBM-MOI}
\end{eqnarray} 
where $\omega$ and $\hat L_{x}$ stand for the cranking frequency and the
$x$-component of the boson angular momentum operator, respectively. 
 
While ${\cal J}_{\rm IBM}$ has six parameters $\epsilon$, $\kappa$, 
$\chi_{\pi}$, $\chi_{\nu}$, $C_{\beta}$ and $\alpha$, all of them but 
$\alpha$ are already fixed by the energy-surface analysis.  
The $\alpha$ value for each nucleus is obtained so that the
${\cal J}_{\rm IBM}$ value at the equilibrium point, where the boson 
energy surface $E_{\rm IBM}(\beta_{B},\gamma_{B})$ is minimal, becomes
identical to the ${\cal J}_{\rm TV}$ value at its corresponding energy minimum. 

The values of all derived IBM parameters are summarized in Table~\ref{tab:IBMpara}.  
When diagonalizing the Hamiltonian in Eq.~(\ref{eq:bh}), the $\epsilon$
parameter is shifted by $\Delta\epsilon=6\alpha$. 
The $\epsilon$ value listed in Table~\ref{tab:IBMpara} is the one with
this shift. 

The diagonalization of the IBM 
Hamiltonian, which is parametrized by the set of interaction strengths summarized in
Table ~\ref{tab:IBMpara}, generates the energies and the wave functions of
the excited states. 
Diagonalization is performed in the boson $M$-scheme basis, where 
$M$ denotes the $z$-component of the boson angular momentum operator. 
With the eigenvectors of the Hamiltonian $\hat H_{\rm IBM}$, the $B$(E2) value is calculated:  
\begin{eqnarray}
 B({\rm E2};L\rightarrow L^{\prime})=\frac{1}{2L+1}|\langle
  L^{\prime}||\hat T^{\rm (E2)}||L\rangle|^2, 
\end{eqnarray}
where $L$ and $L^{\prime}$ are the angular momenta for the initial and
the final states, respectively. 
In the present work the E2 operator is given as 
$\hat T^{\rm (E2)}=e_{\pi}\hat Q_{\pi}+e_{\nu}\hat Q_{\nu}$, where 
$\hat Q_{\rho}$ coincides with the quadrupole operator in
Eq.~(\ref{eq:bh}), and thus the same values of the $\chi_{\pi}$ and
$\chi_{\nu}$ parameters as those listed in
Table~\ref{tab:IBMpara} are used in calculating the $B$(E2)
values (so-called consistent-Q formalism (cf. \cite{IBM})). 
The boson effective charges for protons and neutrons 
are taken the same, namely $e_{\pi}=e_{\nu}$. 

\section{Results and discussion \label{sec:results}}

\subsection{Energy Surfaces}

\begin{figure*}[ctb!]
\begin{center}
\includegraphics[width=16.0cm]{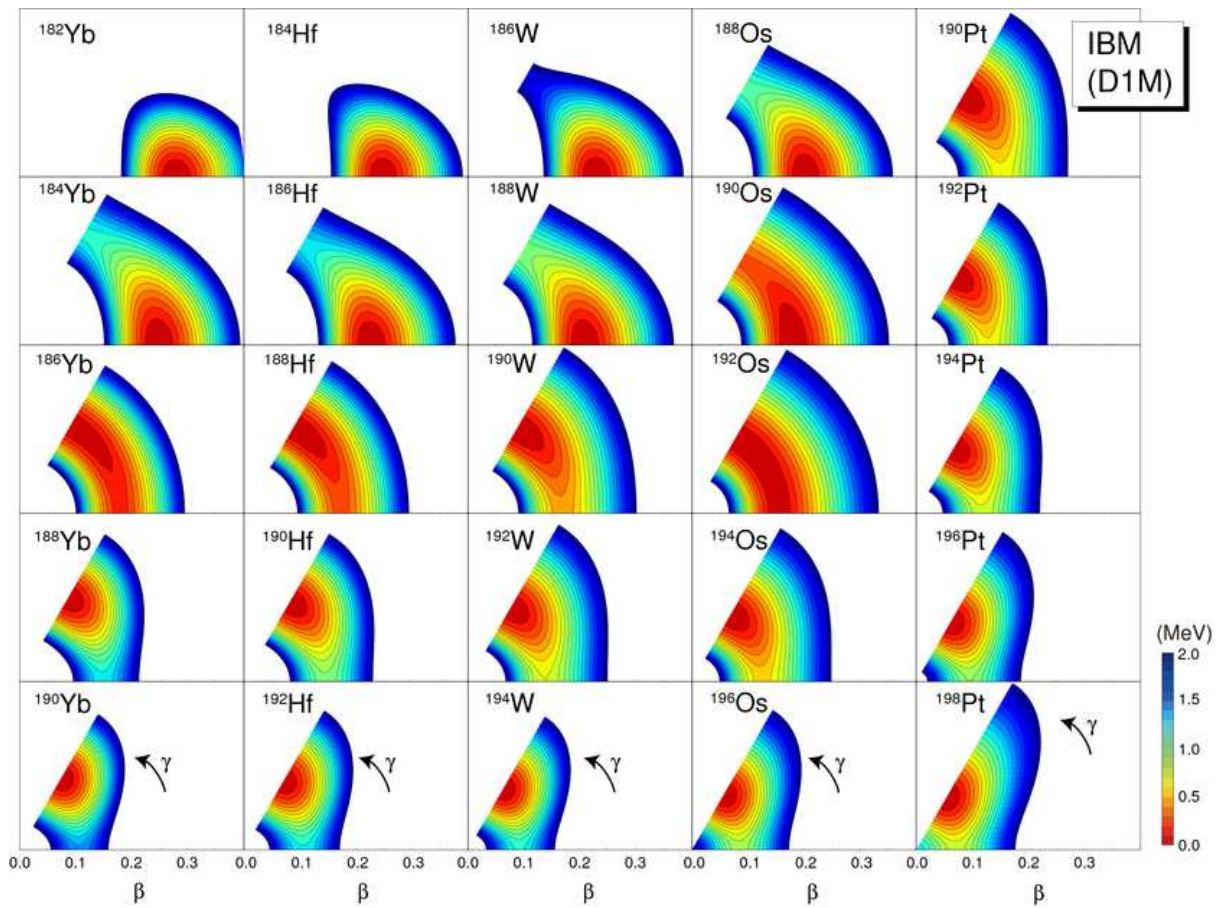}
\caption{(Color online) The IBM energy surfaces for the considered Yb,
 Hf, W, Os and Pt isotopes with $N=112-120$, obtained by the mapping
 from the Gogny-D1M energy surface, depicted within $0\leqslant\beta\leqslant 0.4$ and
$0^{\circ}\leqslant\gamma\leqslant 60^{\circ}$ up to 2 MeV excitation 
from the minimum. Contour spacing is 100 keV. }
\label{fig:ibm-pes}
\end{center}
\end{figure*}

Figure~\ref{fig:ibm-pes} shows the mapped IBM energy surfaces for Yb,
Hf, W, Os and Pt isotopes with $112\leqslant N\leqslant 120$. 
Each energy surface is plotted in terms of
$\beta (=\beta_{B}/C_{\beta})$ and $\gamma(=\gamma_{B})$ up to 2 MeV 
from its absolute minimum,
since most of the quadrupole collective states are within this range. 
Note that the IBM energy surfaces for $N=110$ and 122 are not drawn as
they are similar to those for $N=112$ and 120 nuclei, respectively. 
The Gogny-D1M energy surfaces are not shown as they do not differ 
substantially from the ones depicted in \cite{gradient-2} with Gogny-D1S.

For all the isotopes but the Pt ones, the energy minimum shifts from 
the prolate ($\gamma=0^{\circ}$) to the oblate ($\gamma=60^{\circ}$) 
sides as the number of neutrons increases, passing through the most 
notable $\gamma$-soft nuclei with $N\approx 116$. The derived 
$\chi_{\pi}$ and $\chi_{\nu}$ values for many $N=116$ isotones then 
satisfy $\chi_{\pi}+\chi_{\nu}\approx 0$, as summarized in 
Table~\ref{tab:IBMpara}. This choice of the $\chi$ parameters is at the 
origin of the almost totally flat topology of the energy surface in 
the IBM-2, as seen for example in $^{192}$Os nucleus in 
Fig.~\ref{fig:ibm-pes}. The change in the topology of the energy surface is 
an evidence of prolate-to-oblate shape/phase transition, which 
becomes sharper for smaller $Z$. The Gogny-D1S energy surfaces 
reported in \cite{gradient-2,GognyIBMPt} were somewhat steeper in 
both $\beta$ and $\gamma$ directions than the present Gogny-D1M ones. 

A difference is apparent between the energy surfaces of the Pt 
isotopes and those of the others. For the Pt isotopes, the variation 
of the energy surface takes place much moderately. Such slow 
structural transition in Pt isotopes was also observed in the case 
of the D1S functional \cite{RaynerPt,GognyIBMPt}. While a certain 
quantitative difference is observed between the two Gogny 
functional results, the conclusion does not change.

It should be noted that the Gogny-HFB calculation suggested shallow 
triaxial wells for the transitional, $N=116$ Os and W nuclei 
\cite{gradient-2}. In contrast, the mapped IBM energy surfaces in Fig.~ 
\ref{fig:ibm-pes} are flat in the $\gamma$ direction, as the only 
$\gamma$-dependent term of the bosonic energy surface is 
proportional to $\cos{3\gamma}$. This is the case as long as the 
boson Hamiltonian contains up to two-body interactions. Only when a 
three-body (so-called cubic) term is considered, a stable minimum at 
a $\gamma$ value different from $\gamma=0$ and 60 degrees is obtained
\cite{Heyde1984cubic,Casten-cubic}. 


\subsection{Correlation energies \label{sec:correlation}}

\begin{figure}[ctb!]
\begin{center}
\includegraphics[width=7.0cm]{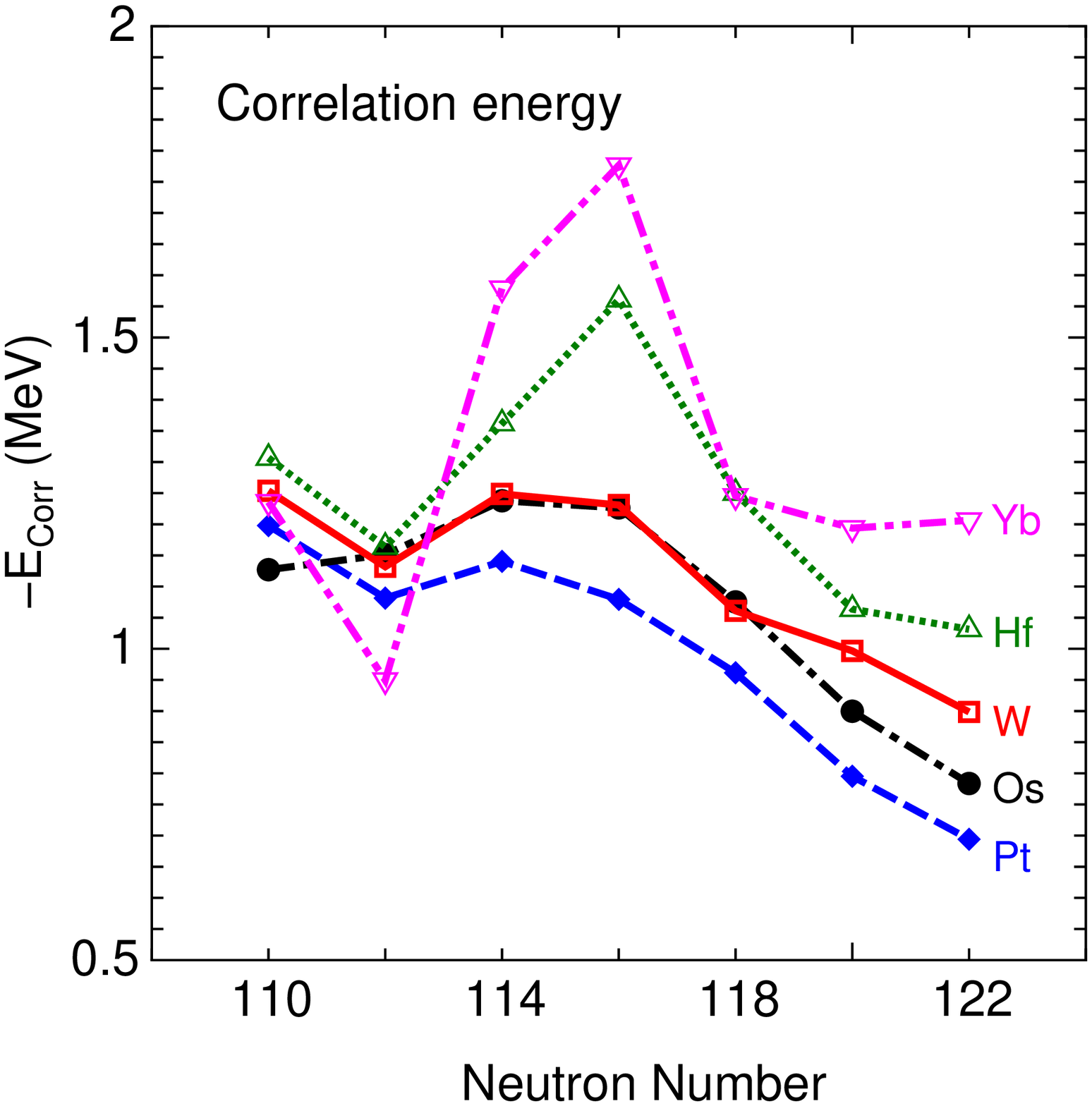}
\caption{(Color online) Correlation energy $E_{\rm Corr}$ defined in
 Eq.~(\ref{eq:corr}) for Yb, Hf, W, Os and Pt isotopes. }  
\label{fig:GE}
\end{center}
\end{figure}

We next discuss a signature for a shape transition from a simple 
perspective. To do this we consider the following quantity that will 
be called {\em correlation energy} hereafter, which was already 
introduced in Ref.~\cite{nsofull}: 
\begin{eqnarray}
 E_{\rm Corr} = E_{\rm IBM}(0^{+}_{1}) - \langle\hat H_{\rm IBM}\rangle_{\rm min}, 
\label{eq:corr}
\end{eqnarray}
where the first term $E_{\rm IBM}(0^{+}_{1})$ is the eigenenergy of the IBM
Hamiltonian, Eq.~(\ref{eq:bh}), for the $L^{\pi}=0^{+}$ ground state, and the second term 
$\langle\hat H_{\rm IBM}\rangle_{\rm min}$ denotes the minimum value of 
the IBM energy surface, that is obtained by the variation with respect
to $\beta$ and $\gamma$. 

In the self-consistent mean-field calculation with a given EDF (e.g.,
Ref.~\cite{Bender2006correlation}), the quantum-mechanical effect can be extracted
by comparing the minimum value of the total energy surface of the mean
field with the $L^{\pi}=0^{+}$ eigenenergy resulting from the 
restoration of the broken symmetries and the configuration mixing. 
For calculations of correlation energies by mapping the EDF theory into
shell model like interactions, including quadrupole and pairing
correlations, the reader is referred to \cite{Rayner-Bertsch}.  


In the present study, all correlation effects can be included by the 
diagonalization of the boson Hamiltonian, and the energies and the 
wave functions of the states with good angular momentum and particle 
number can be generated. Thus, the quantity defined in Eq.~(
\ref{eq:corr}) contains correlation energies coming from symmetry
restoration and configuration mixing and is similar to the equivalent
quantity discussed in GCM studies.

The behavior of $E_{\rm Corr}$ with neutron number correlates well 
with the underlying shape transition. Figure~\ref{fig:GE} shows that 
for each considered isotopic chain the correlation energy is maximal 
in magnitude at the neutron number $N\sim 116$, which corresponds to 
the transition point of the prolate-to-oblate shape transition, and 
decreases as the neutron shell closure $N=126$ is approached. This 
is consistent with the overall systematic trend of the underlying 
energy surface in Fig.~\ref{fig:ibm-pes}. These features have been 
recognized in the GCM studies (e.g., in \cite{Yao2011GCM2}) also. 
For the Pt isotopes, the magnitude of $E_{\rm Corr}$ decreases with 
$N$, indicating that a clear transition is not expected for these 
nuclei.

Compared with the analysis by the GCM configuration mixing using 
e.g., a Skyrme functional \cite{Bender2006correlation} for the same 
mass region as considered here, the magnitude of the present 
correlation energy $E_{\rm Corr}$ is rather small, whereas the 
qualitative features mentioned above do not contradict the GCM 
results. 

In comparison to some rare-earth nuclei such as 
Nd-Sm-Gd isotopes, where a distinct first-order shape transition is 
observed \cite{Cejnar_Jolie_Casten_review}, the shape transition 
occurs rather moderately in the considered mass region. Thus, 
contrary to $E_{\rm Corr}$ in Fig.~\ref{fig:GE}, any drastic change 
with nucleon number is not expected in some other quantities in the 
ground state, like two-nucleon separation energies.

\subsection{Moments of inertia}

Based on the analysis in Sec.~\ref{sec:correlation}, we discuss to 
what extent the moment of inertia is affected by the configuration 
mixing due to the diagonalization of IBM Hamiltonian. The effect is 
most nicely illustrated in the W isotopes, for which relatively many 
experimental spectroscopic data are available. 

\begin{figure}[ctb!]
\begin{center}
\includegraphics[width=7.0cm]{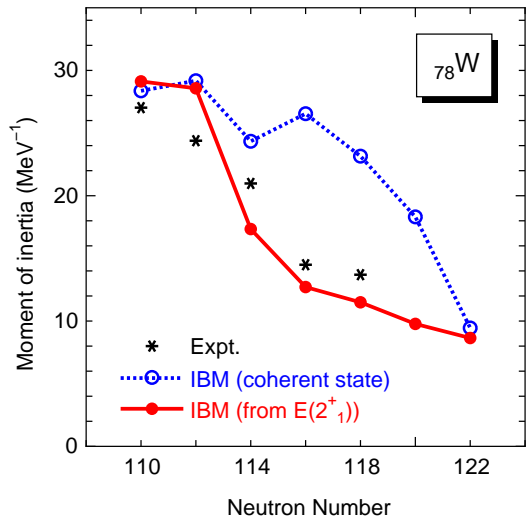}
\caption{(Color online) Moments of inertia of W isotopes, computed by the
 cranking formula for the coherent state, by the the rotor formula $L(L+1)$ using the
 $2^{+}_{1}$ eigenenergies of the IBM and of the experimental
 $2^{+}_{1}$ excitation energies \cite{data} }  
\label{fig:moi}
\end{center}
\end{figure}

We show in Fig.~\ref{fig:moi} the moments of inertia of W isotopes, 
calculated by the cranking formula for the coherent state ${\cal 
J}_{\rm IBM}$ in Eq.~(\ref{eq:IBM-MOI}) and those taken from the 
$2^{+}_{1}$ eigenenergies of the IBM and the experimental $2^{+}_{1}$
excitation energies \cite{data} using the rotor formula $L(L+1)$. 
Note that the cranking moment of inertia of the IBM is, due to the 
correction by the LL term, set identical to the TV moment of 
inertia. Thus the TV moment of inertia is not depicted in Fig.~\ref{fig:moi}. 

The experimental moment of inertia decreases with $N$ and the slope 
of this decrease appears to change at $N=116$. This change suggests 
a gradual shape transition. The moment of inertia of the IBM 
intrinsic state, in contrast, decreases smoothly with the exception 
of the kink at $N=114$. Perhaps such a kink reflects a detailed 
shell structure irrelevant to the present work. However, the kink is 
eliminated in the moment of inertia after diagonalization, which 
falls on the same systematics as the experimental data. 

It appears that, from Fig.~\ref{fig:moi}, the cranking moment of 
inertia still works for the nuclei $N=$110 and 112, for which one 
cannot see any difference from the moment of inertia taken from the 
IBM eigenenergies. In the transitional region of $114\leqslant 
N\leqslant 118$, where according to Fig.~\ref{fig:GE} a large amount 
of correlation energy should be involved, however, the moment of 
inertia of the intrinsic state is far from sufficient and 
configuration mixing by the diagonalization of Hamiltonian becomes 
crucial for the description of the experimental trend. 

\subsection{Excited states \label{sec:level}}

\begin{figure*}[ctb!]
\begin{center}
\begin{tabular}{ccc}
\includegraphics[width=5.6cm]{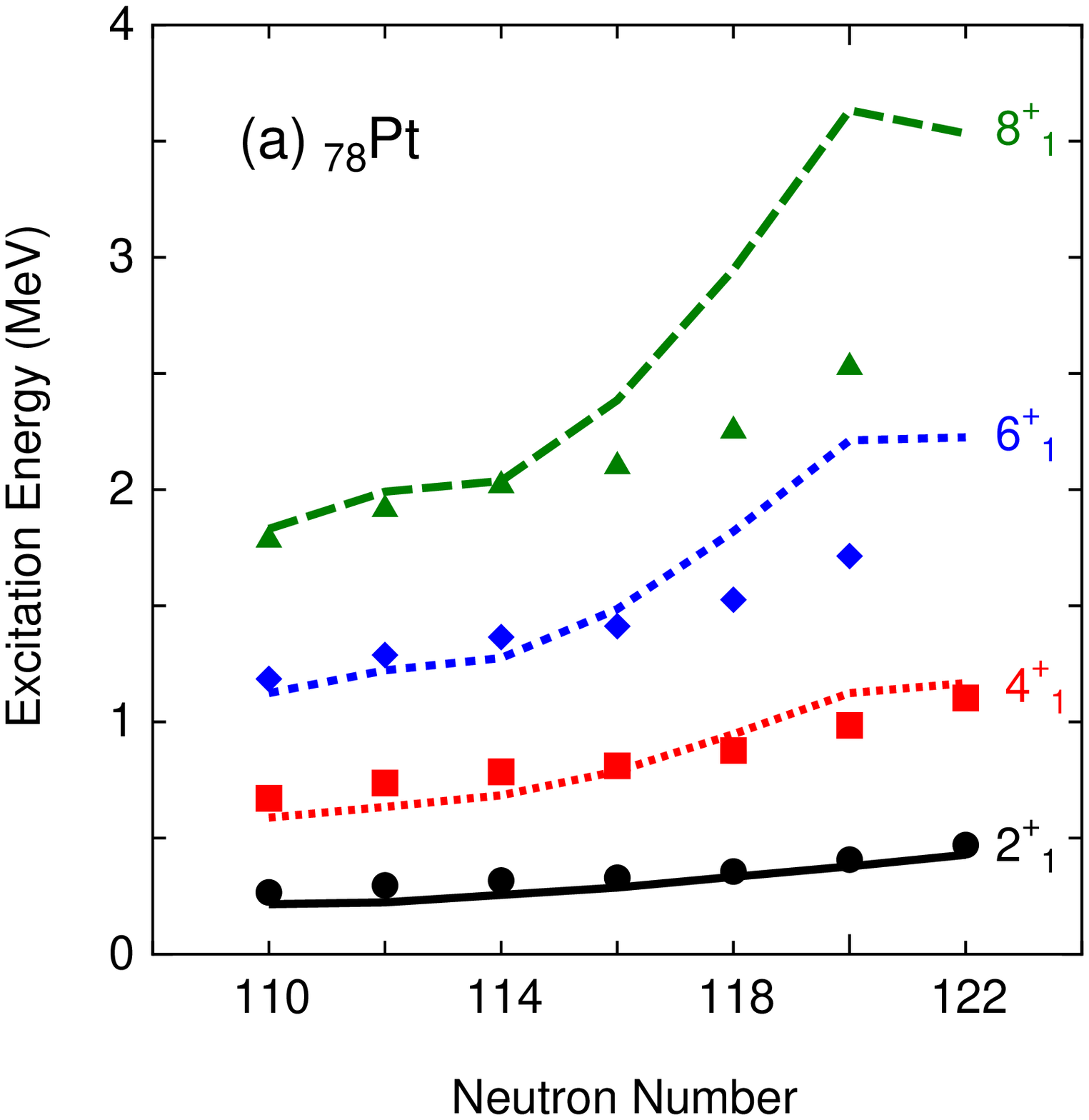} &
\includegraphics[width=5.6cm]{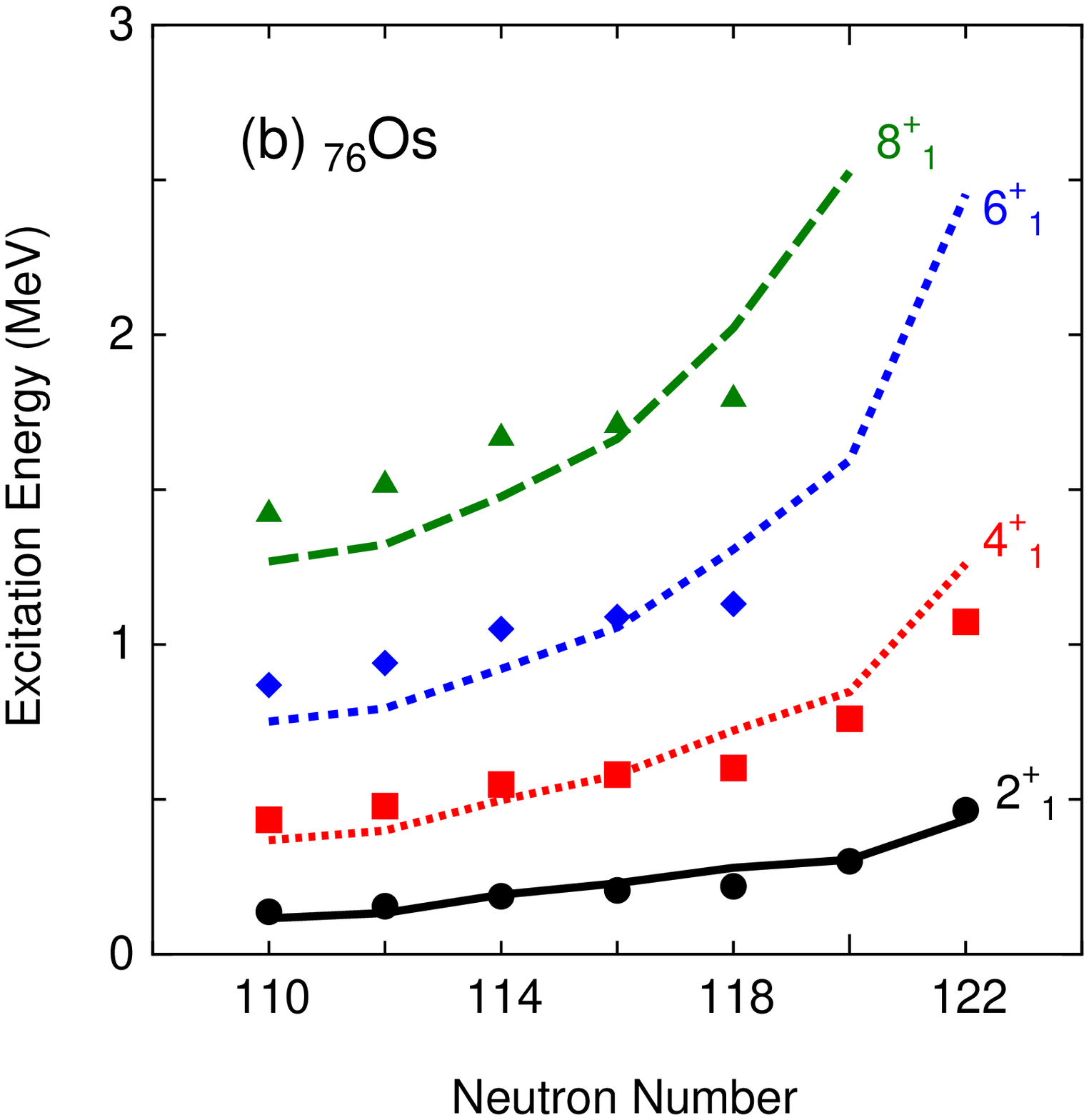} &
\includegraphics[width=5.6cm]{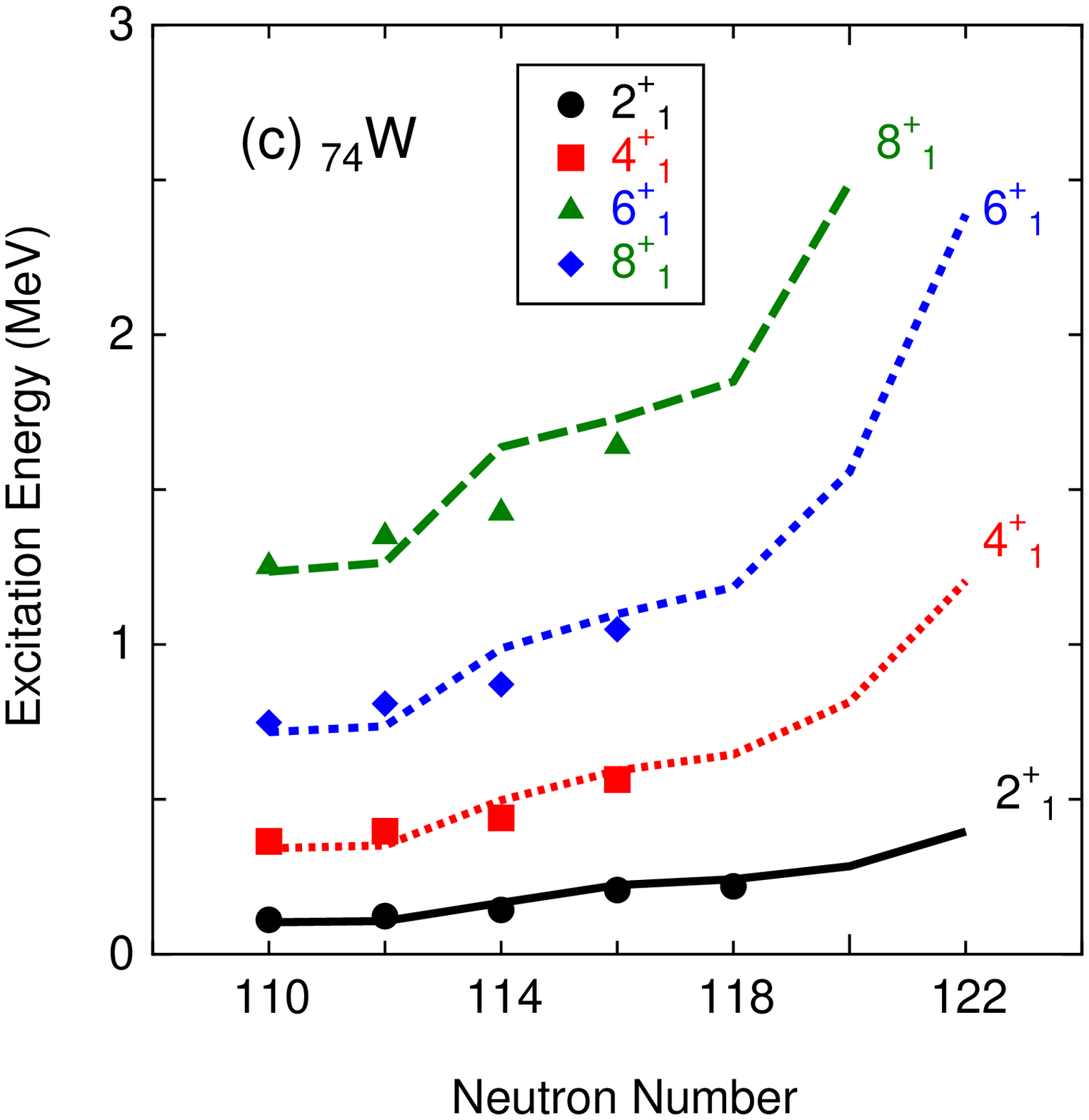} 
\end{tabular}
\end{center}
\begin{center}
\begin{tabular}{cc}
\includegraphics[width=5.6cm]{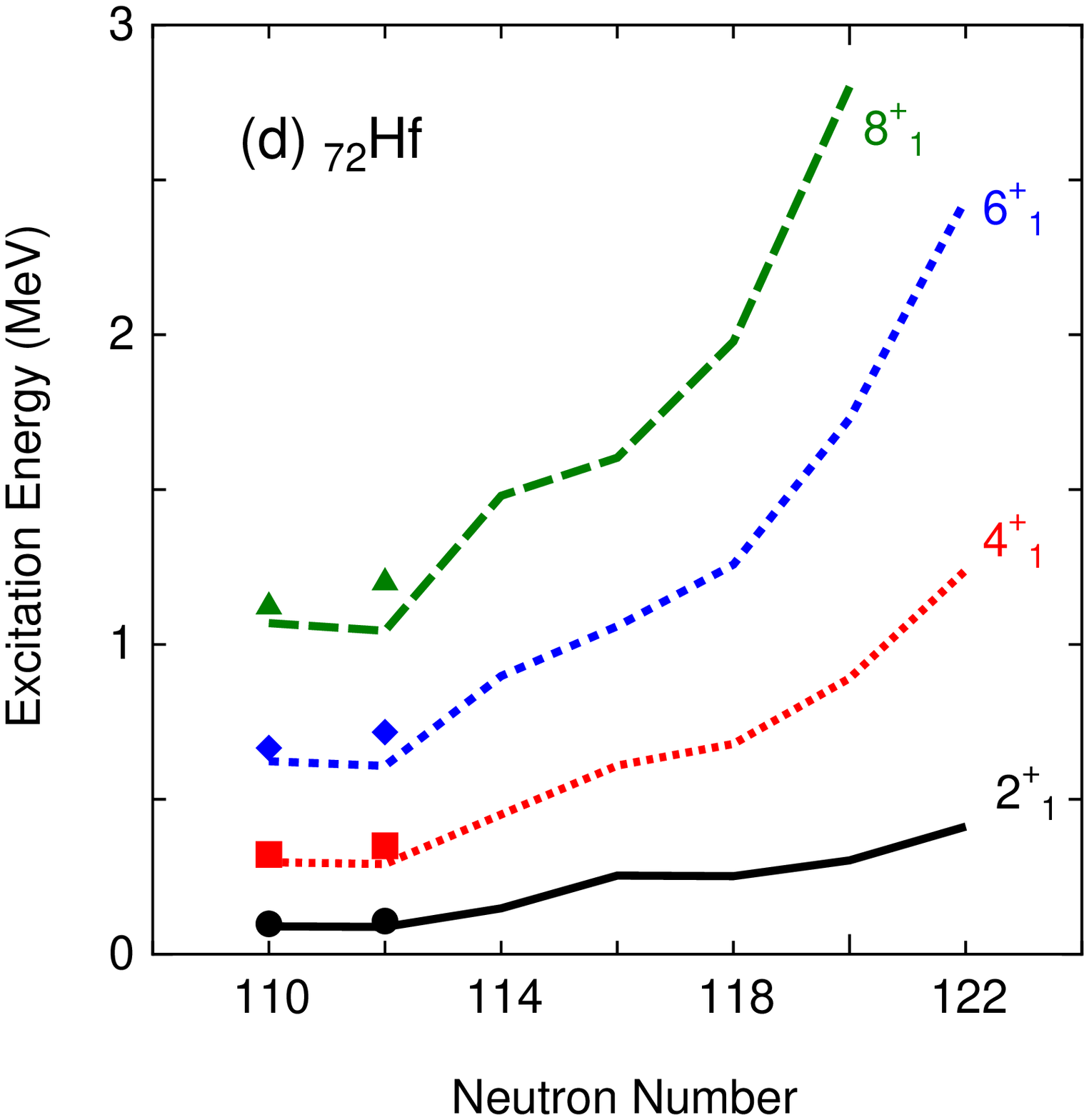} &
\includegraphics[width=5.6cm]{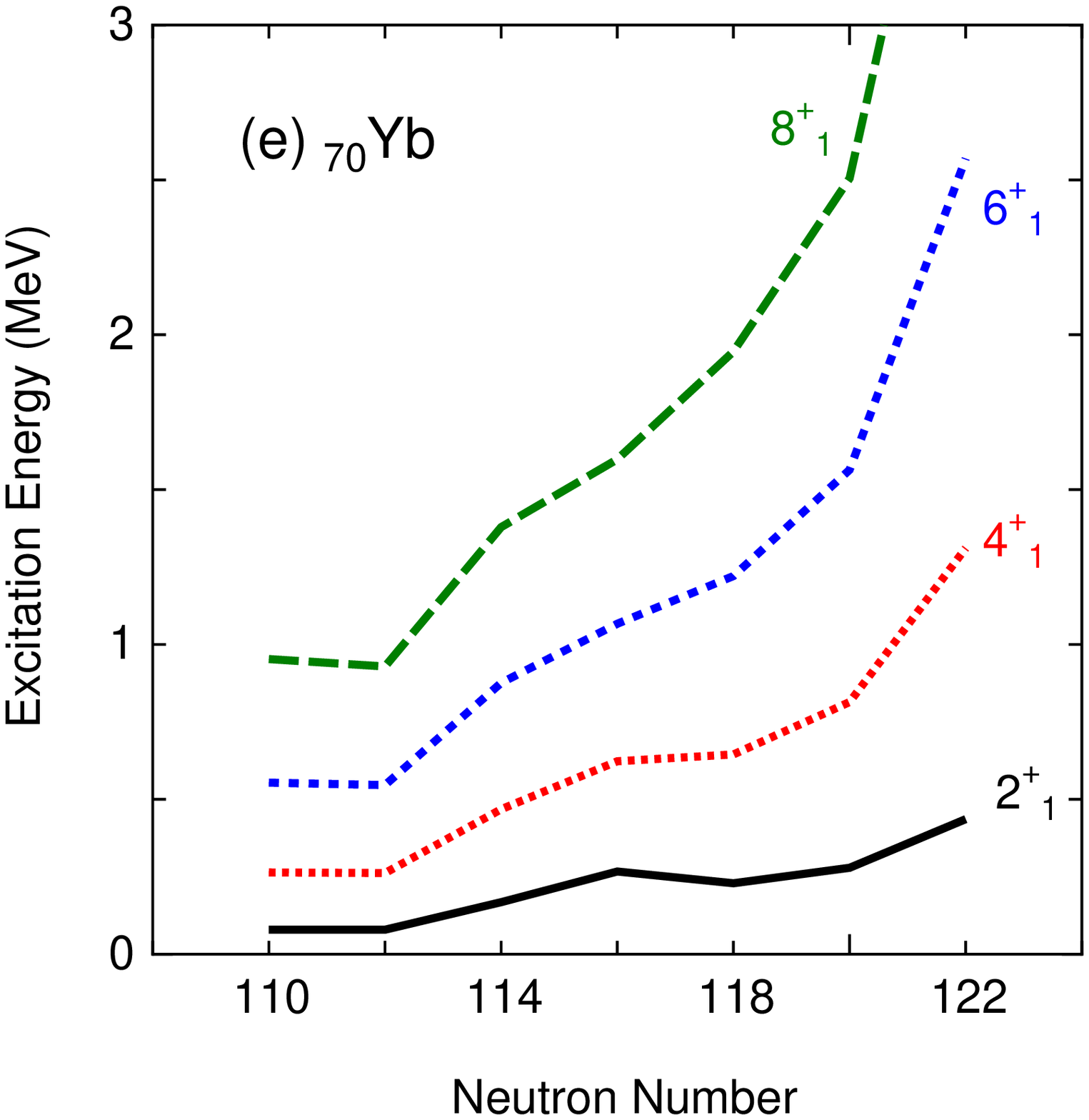} 
\end{tabular}
\caption{(Color online) 
Theoretical (curves) and experimental \cite{data,Alk09,Reg08} (symbols)
 low-lying spectra of Yb, Hf, W, Os, and Pt 
 isotopes with $110\leqslant N\leqslant 122$ for the $2^{+}_{1}$,
 $4^{+}_{1}$, $6^{+}_{1}$ and $8^{+}_{1}$ states. 
Symbols for the experimental levels are defined in the panel (c). } 
\label{fig:level1}
\end{center}
\end{figure*}

\begin{figure*}[ctb!]
\begin{center}
\begin{tabular}{ccc}
\includegraphics[width=5.6cm]{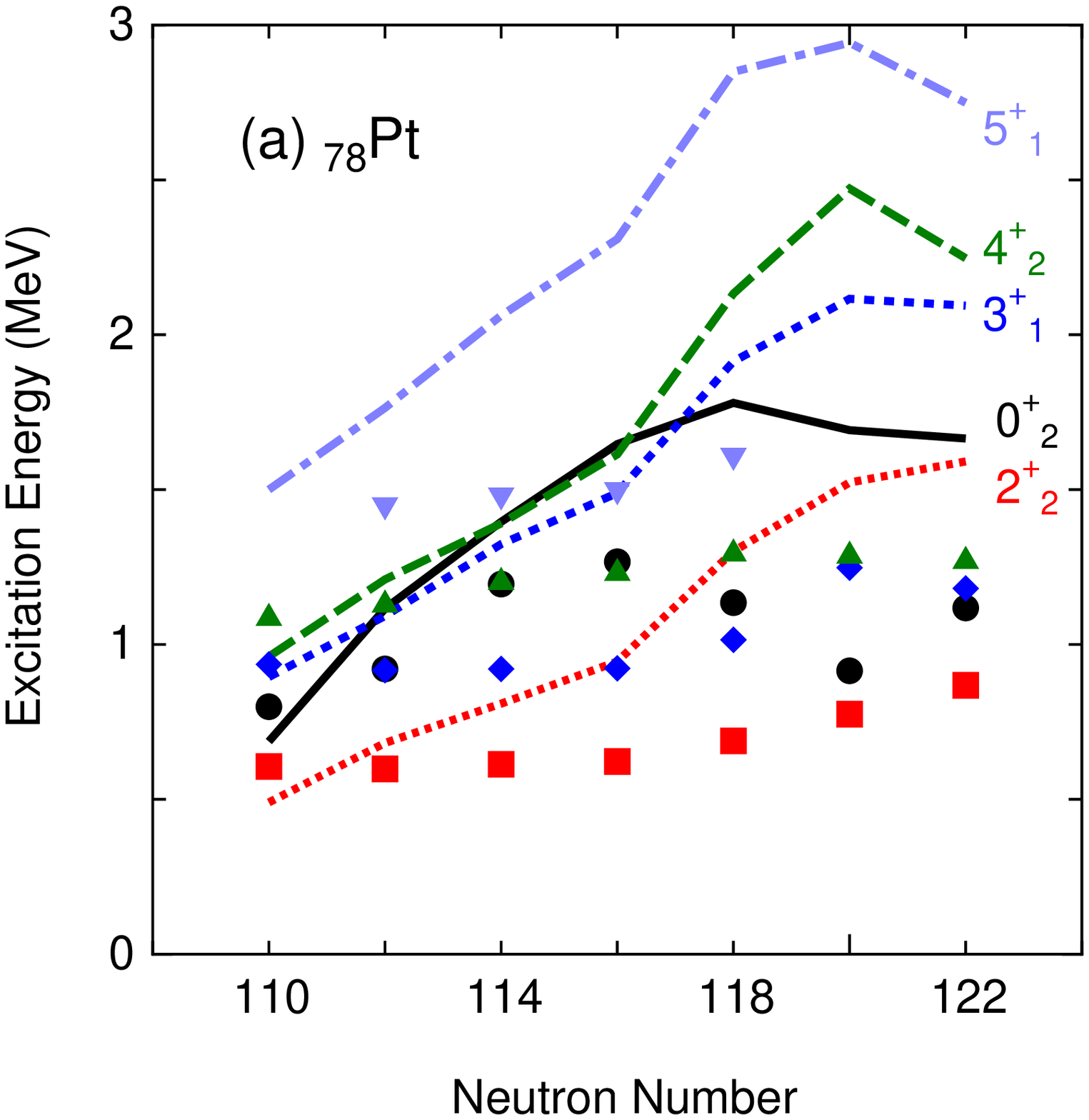} &
\includegraphics[width=5.6cm]{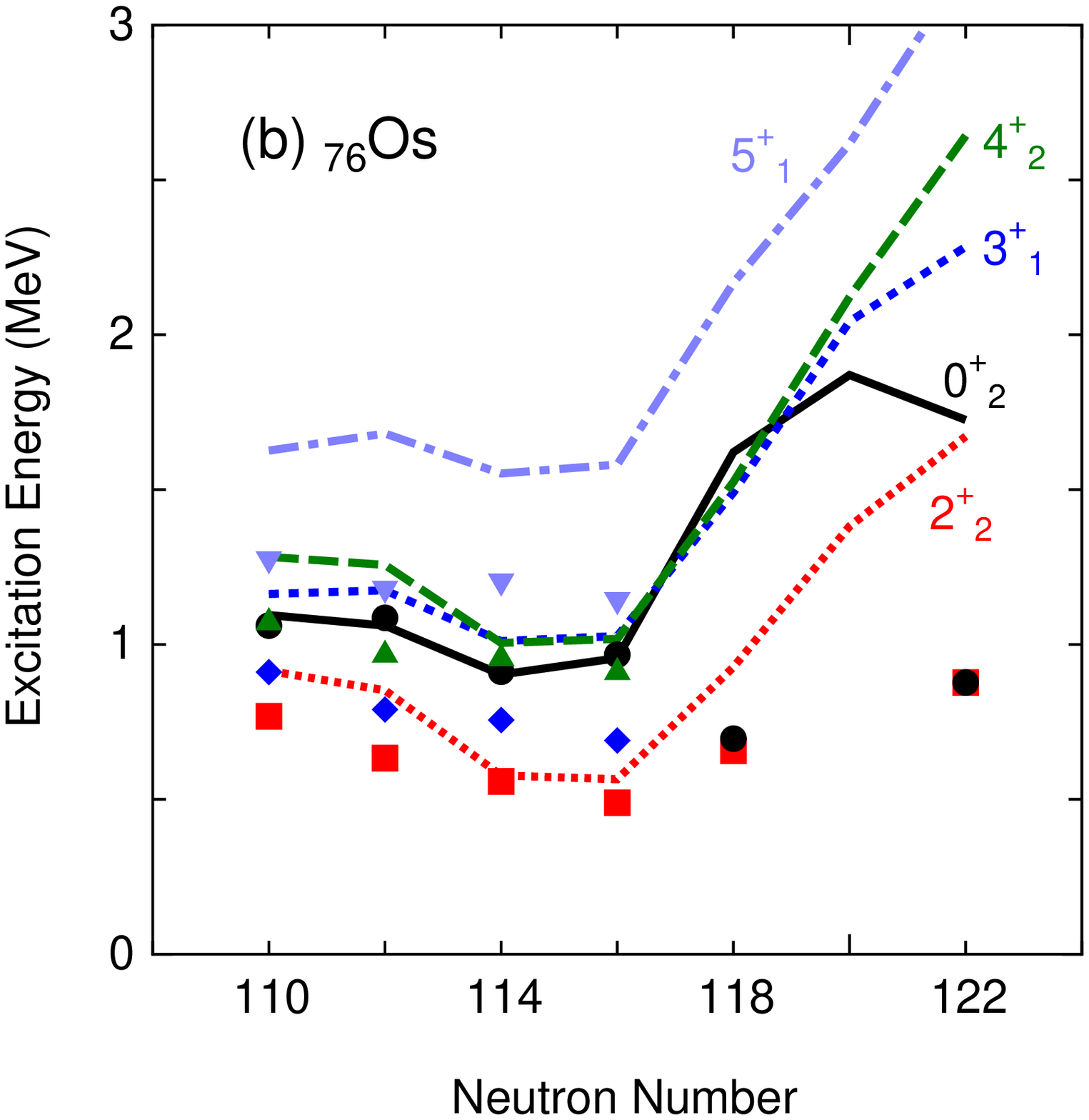} &
\includegraphics[width=5.6cm]{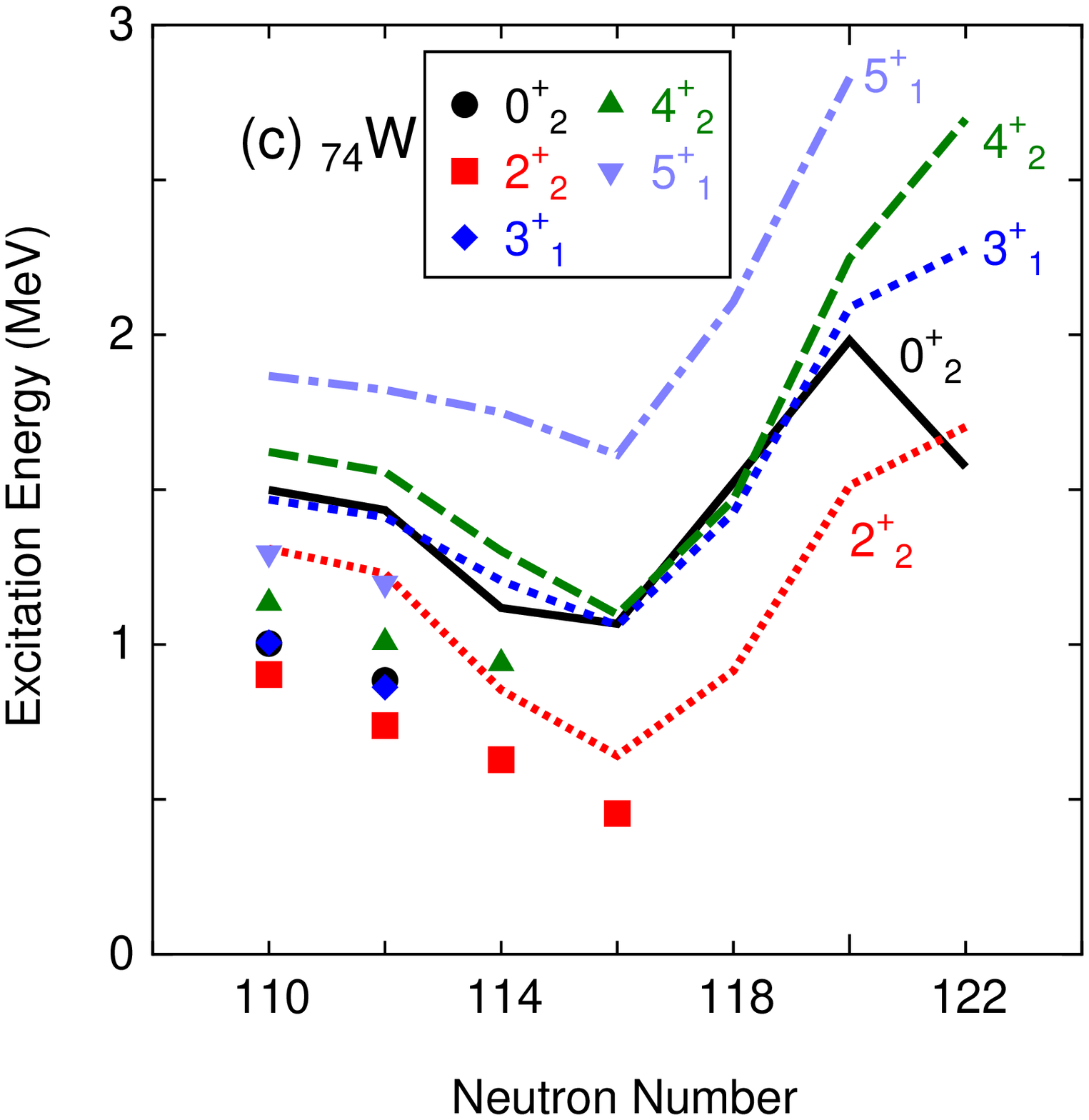} 
\end{tabular}
\end{center}
\begin{center}
\begin{tabular}{cc}
\includegraphics[width=5.6cm]{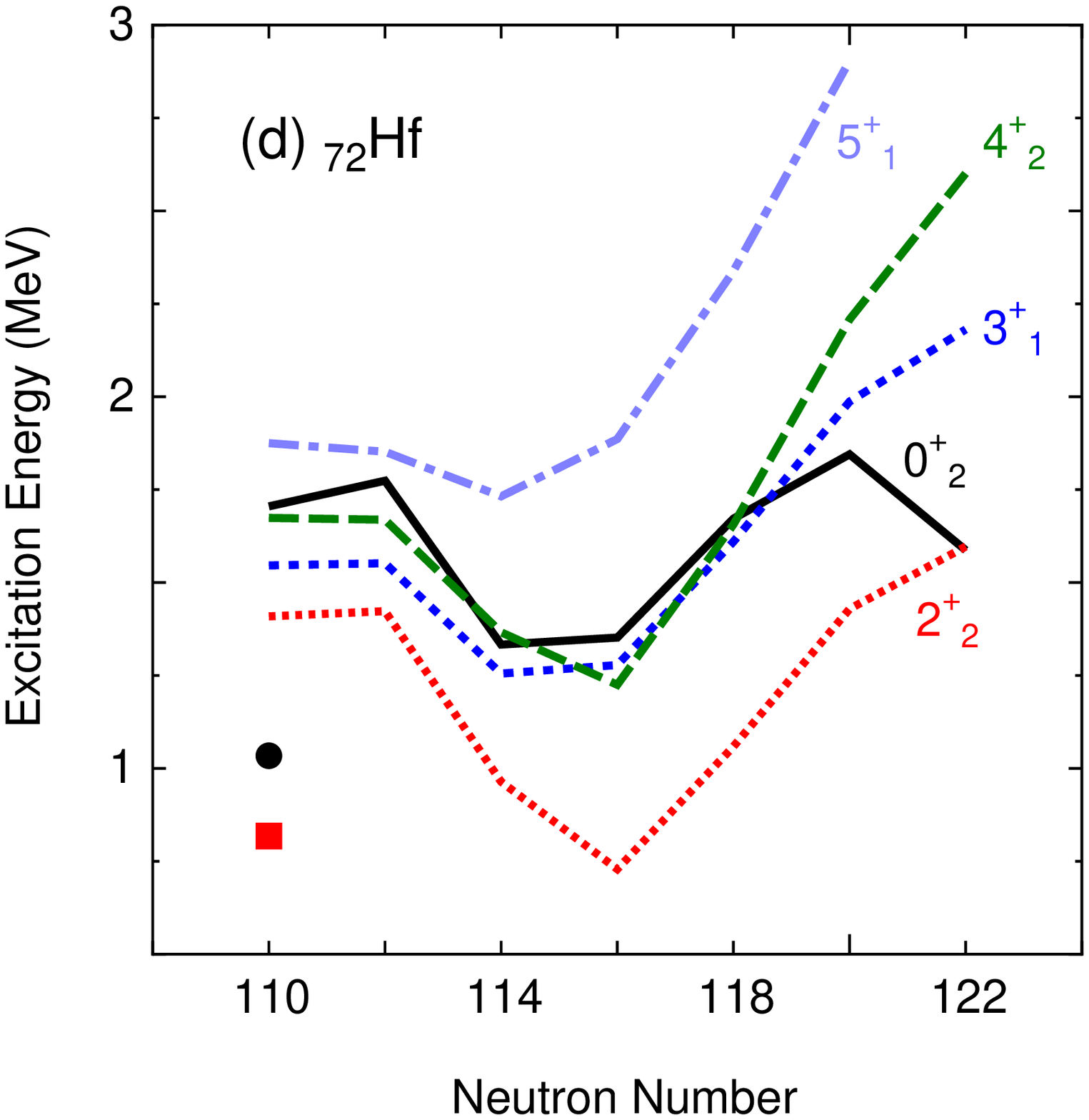} &
\includegraphics[width=5.6cm]{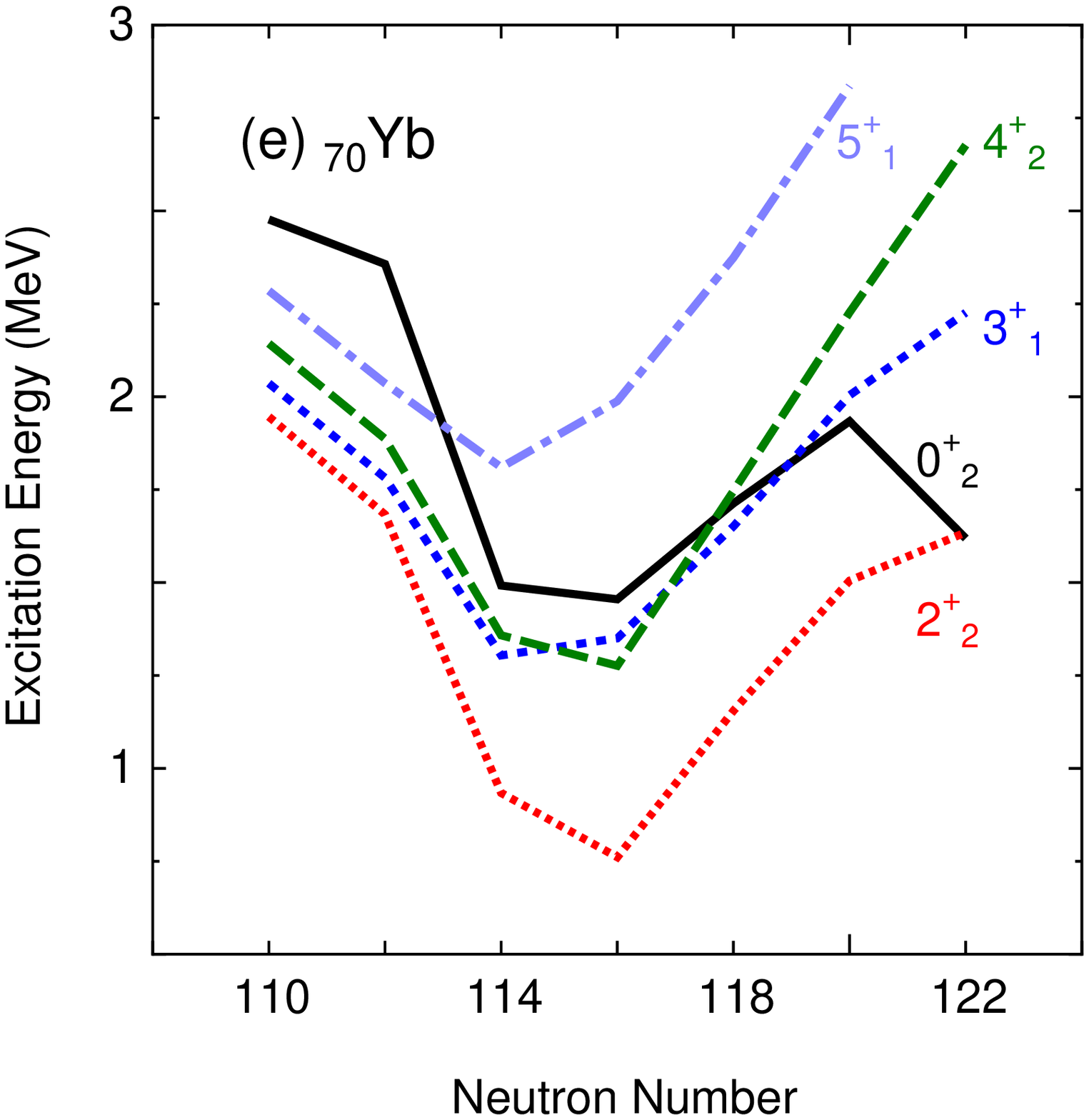} 
\end{tabular}
\caption{(Color online) 
Same as Fig.~\ref{fig:level1}, but for the $0^{+}_{2}$,
 $2^{+}_{2}$, $3^{+}_{1}$, $4^{+}_{2}$ and $5^{+}_{1}$ states.  } 
\label{fig:level2}
\end{center}
\end{figure*}

We now discuss in Figs.~\ref{fig:level1} and \ref{fig:level2} the 
low-lying states for the considered isotopic chains.

Experimentally \cite{data,Alk09,Reg08}, the excitation energies of 
the ground-state band shown in Fig.~\ref{fig:level1}, namely the 
$2^{+}_{1}$, $4^{+}_{1}$, $6^{+}_{1}$ and $8^{+}_{1}$ yrast states, 
increase as the neutron shell closure $N=126$ is approached. The 
increase of these yrast levels with neutron number $N$ becomes more 
rapid with smaller $Z$, when departing from the proton shell closure 
$Z=82$. The present results follow the overall experimental isotopic 
trend for those nuclei. For Pt, Os and W isotopes, the same 
systematics have been observed with the Gogny-D1S 
functional \cite{GognyIBMPt,IBMwos}. 

The LL term has a remarkable influence on the ground-state band at 
the quantitative level. Without this term, the experimental yrast 
spectra would not be reproduced with that precision. This is particularly 
the case with lighter W (Hf) isotopes with $N=110$ and 112, which 
follow the rotor formula $L(L+1)$ with their respective experimental 
ratios being $E_{4^{+}_{1}}/E_{2^{+}_{1}}=$3.27 (3.29) and 3.23 
(3.26) \cite{data}. For these nuclei, the  results shown in Figs.~
\ref{fig:level1}(c) and \ref{fig:level1}(d) compare rather well with 
the experiments.

We now turn to the description of the side-band energies in 
Fig.~\ref{fig:level2}. To begin with, we discuss the excited $0^{+}$ (
$0^{+}_{2}$) state. It is well known that the intruder 
configurations may play a role for mid-shell Pt isotopes, where the 
oblate-prolate shape coexistence is observed \cite{rayner-PRL,Andre}. 
The phenomenological IBM study (see Ref.~\cite{CMIBM}, for instance) 
considers particle-hole excitations across 
the $Z=82$ proton shell. In this kind of work one needs to extend 
the boson model space as to take into account the intruder 
configuration with additional proton bosons, arising from (mainly) 
the $2p$-$2h$ excitation. The normal and the intruder configurations 
are mixed, and the model Hamiltonian should be then diagonalized in 
such enlarged configuration space. The validity of this mixing 
calculation has been discussed extensively \cite{IBM1-Pt-MCZ,IBMCMECQF}, 
and is thus of great interest.
 

The mixing in general becomes more significant when approaching the 
middle of the major shell. In Fig.~\ref{fig:level2}(a), the 
calculated $0^{+}_{2}$ excitation energies for $N\leqslant 116$ Pt 
isotopes, as well as those with Gogny-D1S \cite{GognyIBMPt}, seem to 
compare reasonably well with the data, even without taking into 
account the mixing between normal and intruder states. 
Furthermore, the original HFB energy surfaces for Pt isotopes do not 
exhibit clear coexisting minima. Due to this, the present framework 
cannot fix the parameters for both the normal and the intruder 
configurations as well as those for the operators mixing the two 
configurations. Although such a mixing calculation is a rather 
subtle problem, it is very interesting to study the extent to which 
the intruder configuration plays a role when introduced in the 
present mapping method. 

It was shown experimentally 
\cite{Davidson1994Os,Davidson1999Pt,Kibedi1994Os,Kibedi2001W} that, in 
the non-yrast states of lighter W, Os and Pt nuclei, the band mixing 
could arise more or less from the coexistence of the different 
intrinsic states mentioned above, and makes it rather difficult to 
identify the clear band structure by a model prediction. The 
band-mixing feature should be outside of the model space of bosons 
with low-spin on which the IBM is built, and may be somewhat 
difficult to be reproduced. It is yet not clear whether the similar 
complicated band mixing will be observed in the exotic Yb and Hf 
isotopes. 

The $2^{+}_{2}$ level, which is normally the band-head of the 
$K^{\pi}=2^{+}$ (so-called quasi-$\gamma$) band, is a good test for 
the evolving triaxiality in a given isotopic chain. 
Figure~\ref{fig:level2} shows that the calculated $2^{+}_{2}$ level of the 
$N=116$ nuclei is lowest among each of Yb, Hf, W and Os isotopes. 
Experimental excitation energies keep steady (decrease) in Pt (Os, 
W) isotopes as $N$ increases from 110 to 116. 

In our calculations, the decrease of the energies of the $2^{+}_{2}$, 
$3^{+}_{1}$, $4^{+}_{2}$ and $5^{+}_{1}$ states occurs more 
rapidly for lower $Z$ isotopes, which have a larger number of active 
bosons. Around $N=116$ a change in this tendency occurs and these 
excitation energies increase. This is in agreement with the only 
experimental measurement available in Os isotopes. 


A remarkable difference between the theoretical and the experimental 
quasi-$\gamma$-band structure observed in Pt and Os isotopes is that 
the calculated $3^{+}_{1}$ and the $4^{+}_{2}$ states, and the $5^{+}_{1}$ and 
the $6^{+}_{2}$ states as well,  form doublets, 
which are absent in the data. Since all the states in 
Figs.~\ref{fig:level2}(a) and \ref{fig:level2}(b) except the $0^{+}_{2}$ ones, 
are supposed to be the quasi-$\gamma$ band states, the appearance of 
these doublets points to the emergence of the $\gamma$-unstable 
\cite{gsoft} or O(6) dynamical symmetry \cite{IBM}, in which the spectra 
belonging to the same family of the quantum number $\tau$ are nearly 
degenerated. Since the rigid triaxial rotor model with 
$\gamma=30^{\circ}$ \cite{triaxial} predicts the doublets ($2^{+}$, 
$3^{+}$), ($4^{+}$, $5^{+}$), etc, in the $\gamma$ band, the 
experimental data in Fig.~\ref{fig:level2} for (a) Pt, (b) Os, and 
(c) W isotopes suggest a situation rather in between the $\gamma$
-unstable rotor and the rigid-triaxial rotor pictures. The 
discrepancy of the $\gamma$-band energies occurs probably because 
the IBM energy surface does not show the triaxial minimum which is, 
however, seen in the original HFB energy surface. 

There are several possible effects which may eliminate this 
staggering in the $\gamma$-band spectra and improve the agreement 
with the experiments at the quantitative level. In the present 
paper, however, we do not look into the details of this issue due to 
the large number of additional parameters to be introduced and the 
lack of  experimental data for the Yb and Hf nuclei. 
First, a three-body (cubic) term, which partially breaks O(6) 
symmetry, may correct the deviation. This has been done mainly in 
the IBM-1 \cite{Heyde1984cubic,Casten-cubic}. For the present case 
some type of cubic term appears to be necessary mainly for W, Os and 
Pt nuclei, where the Gogny HFB energy surface exhibits a shallow, 
but  stable triaxial minimum \cite{gradient-2}. While the 
calculated excitation energies of the quasi-$\gamma$ band for Yb and 
Hf in Fig.~\ref{fig:level2}(d,e) look like that of pure O(6) limit 
as well, the validity of this term seems to be marginal in these 
cases. Indeed for the Yb and Hf isotopes the original Gogny-D1M 
energy surface indicates the discrete change of the minimum point 
from the oblate ($\gamma=60^{\circ}$) to the prolate 
($\gamma=0^{\circ}$) sides, similarly to the Gogny-D1S energy surface 
\cite{gradient-2}. 

The second possibility would be to relax the constraint on the 
deformation parameters $\gamma_{\pi}$ and $\gamma_{\nu}$ so that 
they could take different values. As the IBM-2 can be viewed as a 
two-fluid system consisting of proton and neutron bosons, the 
phase-structure analysis would be exploited in the context of the 
coherent-state formalism \cite{Caprio2005QPT}, whereas it is not 
obvious to define a consistent mapping procedure for realistic cases. 

The third would be the inclusion of higher-spin bosons, like the $g$-boson. 
It is not independent of the first possibility involving the cubic term, since
the cubic term can be derived effectively from the renormalization of the $g$
boson into the $sd$-boson sector \cite{Heyde1984cubic}. 
This would, of course, make the problem more complicated.

We now address the problem of why the side-band spectra, 
particularly for Pt in Fig.~\ref{fig:level2}(a) and Os in 
Fig.~\ref{fig:level2}(b) isotopes, are overestimated in the present 
calculation when approaching the $N=126$ shell closure. The direct 
reason would be that the microscopic Gogny energy-surface 
calculation predicts mostly oblate deformations with small 
quadrupole moment but with rather large amount of deformation 
energy characterized by the depth of the potential minimum 
\cite{gradient-2}. Such a topology of the HFB energy surface is not well 
described by the IBM Hamiltonian close to the end of the major shell 
$Z=82$. Nearby the closed shell one has a relatively small number of 
bosons. The deviation of the spectra seems to be due to this limited 
degrees of freedom. The problem on the description of the side-band 
energies was observed in other cases of shape transitions in 
different mass regions \cite{nso,nsofull}, and is still an open 
problem. According to the above argument it may be expected that the 
predicted levels for exotic Yb and Hf isotopes in the vicinity of 
the shell closure $N=126$ might be overestimated.

\begin{figure*}[ctb!]
\begin{center}
\begin{tabular}{cc}
\includegraphics[width=7.0cm]{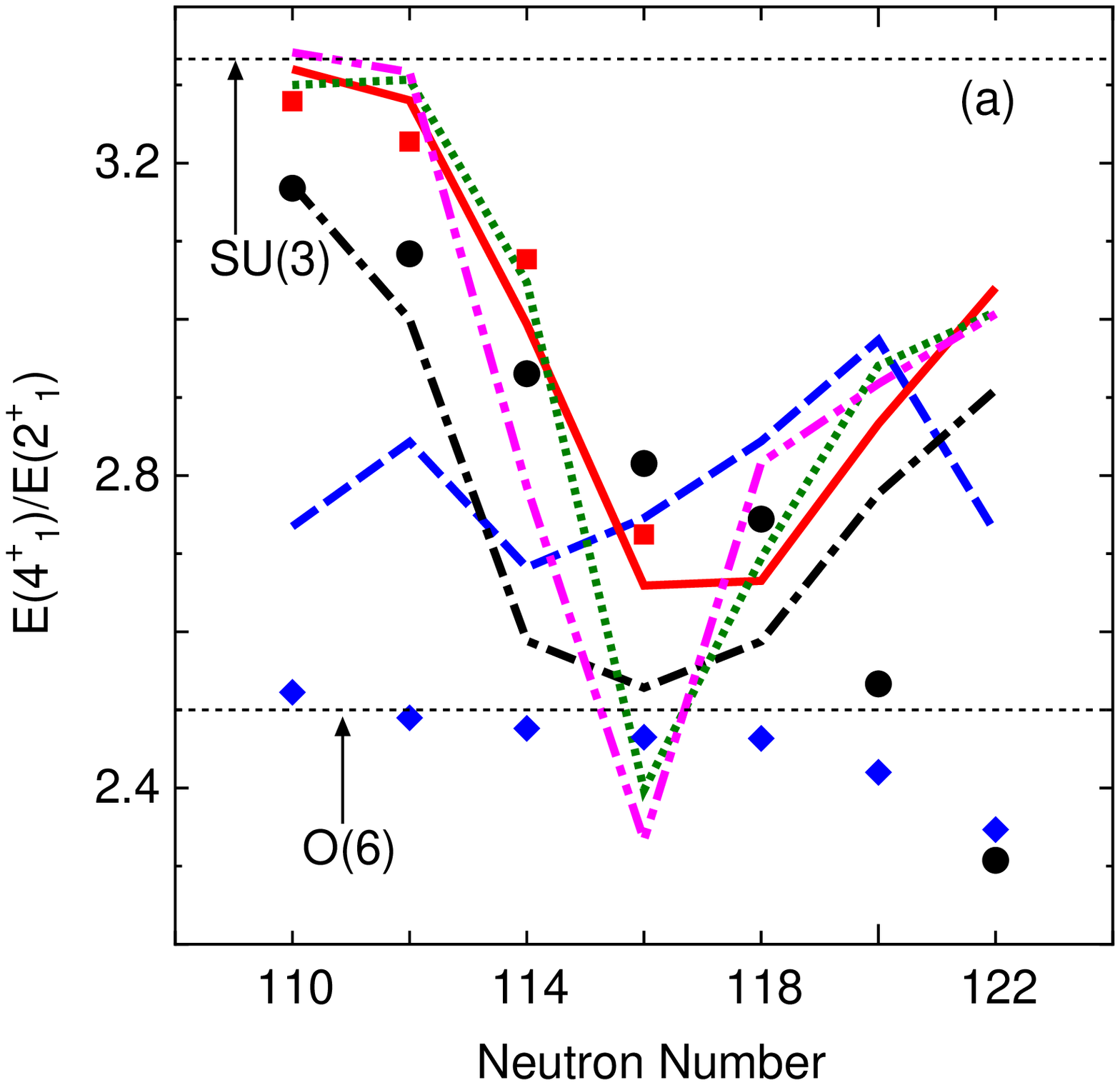} &
\includegraphics[width=7.0cm]{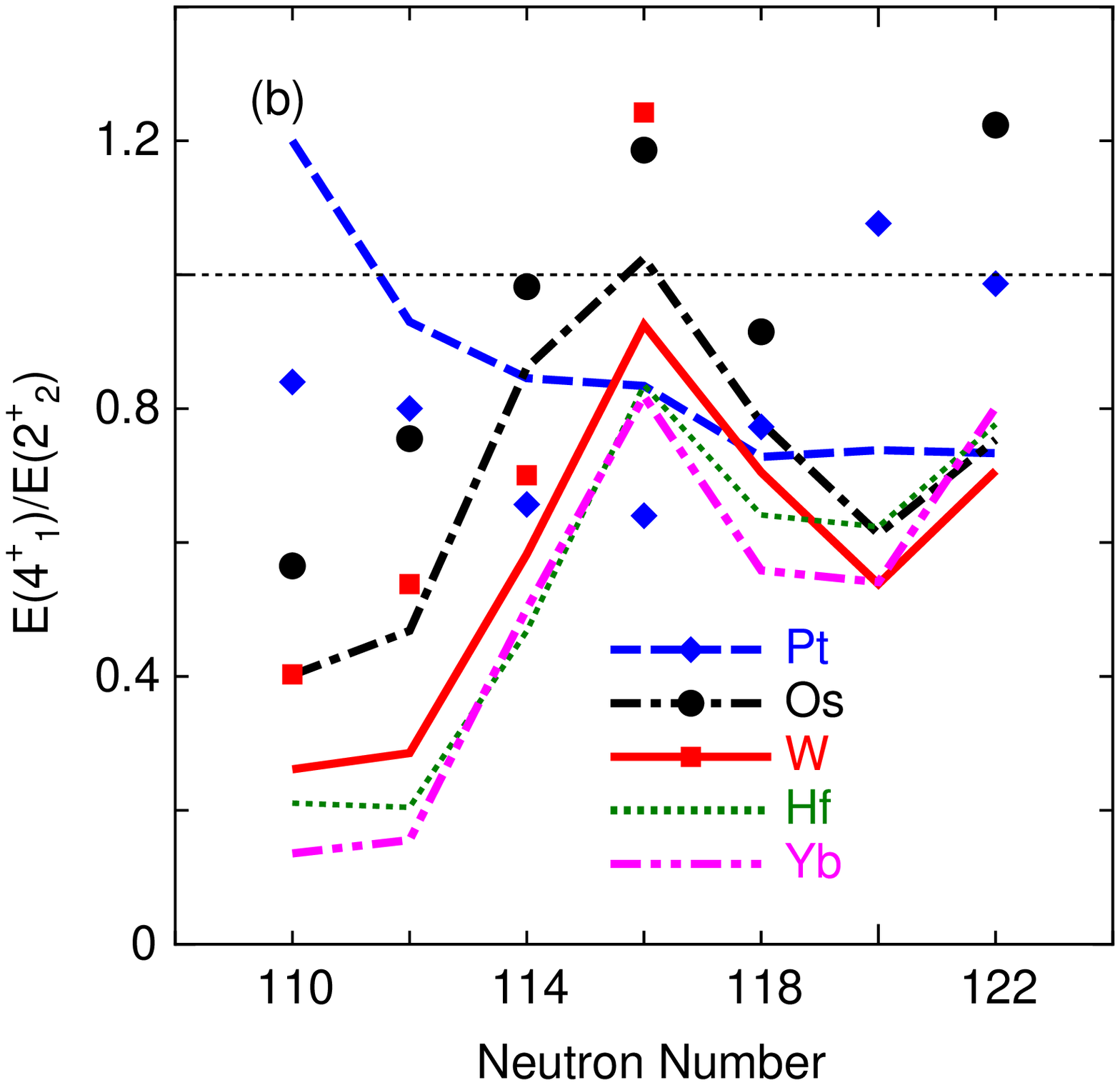}
\end{tabular}
\caption{(Color online) Theoretical (curves) and experimental
 (symbols) \cite{data} energy ratios 
(a) $R_{4/2}=E(4^{+}_{1})/E(2^{+}_{1})$ and 
(b) $R_{4\gamma}=E(4^{+}_{1})/E(2^{+}_{2})$ as functions of $N$.  
Definitions of the theoretical curves and the symbols for the experimental
 data appear in panel (b). }
\label{fig:ratio}
\end{center}
\end{figure*}

To further examine the problem, it is interesting to consider the relevant
energy ratios, as they nicely trace the underlying shape transition. 
Figure~\ref{fig:ratio} depicts the energy ratios (a) $R_{4/2}\equiv E(4^{+}_{1})/E(2^{+}_{1})$ and 
(b) $R_{4\gamma}\equiv E(4^{+}_{1})/E(2^{+}_{2})$ as functions of $N$. 
The ratio $R_{4/2}$ is probably the simplest and best-studied measure for
the evolution of collectivity.  
The ratio $R_{4\gamma}$ presents the location of
the band-head of the quasi-$\gamma$ band $2^{+}_{\gamma}$ ($2^{+}_{2}$)
relative to the $4^{+}_{1}$ excitation energy. 
Since in many $\gamma$-soft nuclei the $2^{+}_{2}$ level lies quite
close to the $4^{+}_{1}$ level, the overall trend of ratio $R_{4\gamma}$
can help to measure the $\gamma$ softness.  

In Fig.~\ref{fig:ratio}(a), the experimental 
$R_{4/2}$ ratios for Os and W isotopes exhibit a gradual decrease as a
function of $N$ from the rotor limit of $R_{4/2}=3.3$ in the vicinity of $N=110$ toward the
O(6) limit of $R_{4/2}=2.5$.
This reflects the transition from the axially deformed rotor to the $\gamma$-unstable shape. 
Also of particular interest is the difference of the $R_{4/2}$ ratio 
between Pt isotopes and the other isotopes. 
The experimental $R_{4/2}$ ratio for all the Pt isotopes studied remains practically 
constant all the way, being close to the O(6) limit of 2.5. 
The present calculation follows the decrease of the experimental $R_{4/2}$ value from
$N=110$ to 116 in Os and W isotopes, while an increase 
is suggested for not only Os and W but Hf and Yb isotopes for
$N\geqslant 118$, contrary to the experimental tendency of Os isotopes. 
The change in the calculated ratio $R_{4/2}$ occurs quite rapidly for Hf and Yb
isotopes in comparison to W and Os isotopes. 
The discrepancy of the tendency for $N\geqslant 118$ for Os nuclei could
be the consequence of the unexpectedly large $\chi_{\pi}$ and 
$\chi_{\nu}$ values with positive sign, as seen in 
Table~\ref{tab:IBMpara}, since the corresponding IBM energy surfaces
exhibit notable oblate deformation. 
The same would hold for explaining the overall deviation in Pt isotopes. 
In this context, to describe all the observed data including those for
$N\geqslant 118$ regime, the triaxial dynamics needs to be correctly
incorporated in the present model. 

The energy ratio $R_{4\gamma}$ is depicted in Fig.~\ref{fig:ratio}(b). 
The experiment shows that in the lighter Pt, Os and W isotopes 
with $N=110$, 112 and 114, the ratio is below unity. 
While for Pt isotopes the experimental ratio $R_{4\gamma}$ remains all the way with values close to unity,
for Os and W isotopes the $\gamma$ softness gradually develops with $N$
as the ratio $R_{4\gamma}$ increases for $110\leqslant N\leqslant 116$ and
overpasses $R_{4\gamma}=1$ at $N=116$.   
The overall trend of this experimental ratio for W and Os isotopes is
reproduced in the present calculation, and the same systematic trend is
predicted for Yb and Hf isotopes. 
For Os, the experimental $R_{4\gamma}$ ratio decreases from $N=116$ to
$118$, which is reproduced by the calculation. 
In the heavier isotopes with $N\geqslant 118$ there is a new tendency
that the calculated ratio shows overall decrease, being much below the 
unity, whereas the experimental
ratio for Os isotopes keeps increasing, being larger than unity. 
The results presented here do not differ much from the case of D1S
functional already studied in \cite{GognyIBMPt,IBMwos}.

\subsection{$B$(E2) systematics}

\begin{figure*}[ctb!]
\begin{center}
\begin{tabular}{cc}
\includegraphics[width=7.0cm]{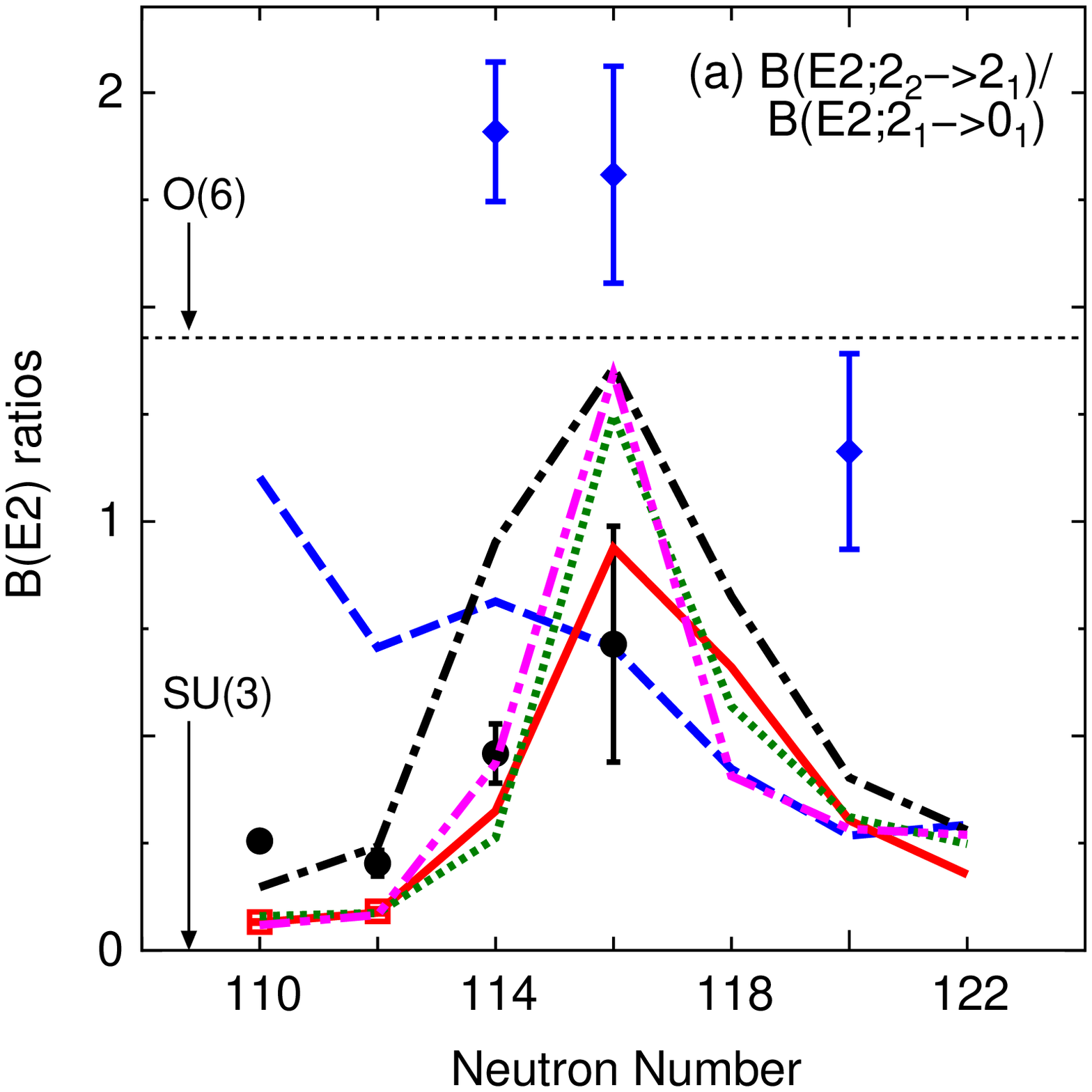} &
\includegraphics[width=7.0cm]{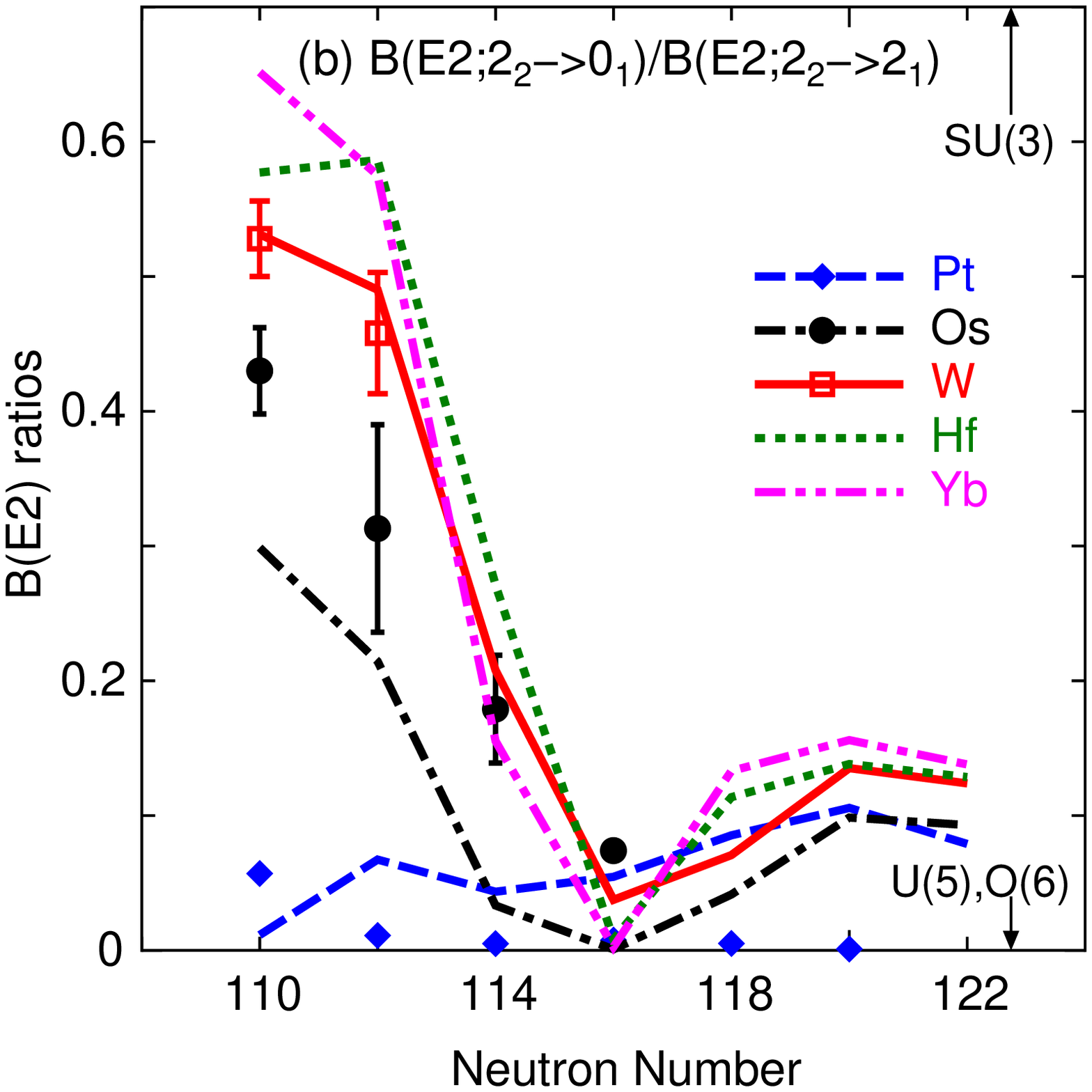}
\end{tabular}
\caption{(Color online) The $B$(E2) ratio (a) 
$B({\rm E2};2_2^+\rightarrow 2_1^+)/B({\rm  E2};2_1^+\rightarrow 0_1^+)$ 
and the branching ratio (b)
$B({\rm E2};2_2^+\rightarrow 0_1^+)/B({\rm E2};2_2^+\rightarrow 2_1^+)$ for relevant low-lying states
of the considered Yb, Hf, W, Os, and Pt isotopes with Gogny-D1M EDF. 
Experimental data for W, Os, and Pt isotopes are taken from
Ref.~\cite{BE2}. 
Definitions of symbols and theoretical curves appear in panel (b). }
\label{fig:BE2}
\end{center}
\end{figure*}

Lastly, we examine the $B$(E2) systematics for a few essential cases
corresponding to the shape transition. 
The $B$(E2) ratios relevant to the band-head of quasi-$\gamma$ band,
$2^{+}_{2}$ state, can be the stringent tests. 

We show in Fig.~\ref{fig:BE2} the ratio 
(a) $B({\rm E2};2_2^+\rightarrow 2_1^+)/B({\rm  E2};2_1^+\rightarrow 0_1^+)$ 
and the branching ratio 
(b) $B({\rm E2};2_2^+\rightarrow 0_1^+)/B({\rm E2};2_2^+\rightarrow 2_1^+)$ 
for the considered isotopes in comparison with the data \cite{BE2note,BE2}. 

The $2_2^+\rightarrow 2_1^+$ E2 transition rate shows a certain 
sensitivity to the neutron number $N$ and thus it is useful as a 
signature of the structural evolution involving the $\gamma$ 
softness. The $B({\rm E2};2_2^+\rightarrow 2_1^+)/B({\rm  
E2};2_1^+\rightarrow 0_1^+)$ ratios for Pt isotopes differ notably 
from those of other isotopes. For Yb, Hf, W, and Os isotopes, the 
calculated ratio is peaked at $N=116$. This confirms that in each of 
these isotopic chains the $N=116$ nucleus is softest in $\gamma$ 
direction. 
On the other hand, for Pt 
isotopes the calculated $B({\rm E2};2_2^+\rightarrow 2_1^+)/B({\rm  
E2};2_1^+\rightarrow 0_1^+)$ value keeps increasing toward $N=110$ 
to approach the O(6) limit, rather than taking a maximum at 
$N=116$. This tendency appears to be consistent with that expected 
from the topology of the HFB energy surface \cite{RaynerPt} and from 
the predicted systematics of the quasi-$\gamma$ band-head in 
Fig.~\ref{fig:level2}(a), which reflects that the $\gamma$ softness persists 
for rather wide region in the Pt isotopic chain. 

When compared with the D1S case \cite{IBMwos}, the 
present D1M result suggests that the ratio $B({\rm 
E2};2_2^+\rightarrow 2_1^+)/B({\rm E2};2_1^+\rightarrow 0_1^+)$ is 
rather sensitive to the isotopic chains. In fact, in 
Fig.~\ref{fig:BE2}(a), the $B({\rm E2};2_2^+\rightarrow 2_1^+)/B({\rm  
E2};2_1^+\rightarrow 0_1^+)$ values below and above $N=116$ appear 
to have a certain $Z$ dependence when the D1M functional is used. 
For instance, the $B({\rm E2};2_2^+\rightarrow 2_1^+)/B({\rm  
E2};2_1^+\rightarrow 0_1^+)$ value for W isotopes is generally far 
from the O(6) limit all the way. It has been noticed in 
Ref.~\cite{IBMwos}, however, that the calculated value of this $B$(E2) ratio 
is practically the same for Os and W isotopes when the D1S 
functional is taken. It would be interesting to see if this
$Z$ dependence is observed experimentally.

The branching ratio $B({\rm E2};2_2^+\rightarrow 0_1^+)/B({\rm 
E2};2_2^+\rightarrow 2_1^+)$ in Fig.~\ref{fig:BE2}(b) also presents 
a clear signature of the structural evolution involving triaxiality. 
For Yb, Hf and W isotopes with $110\leqslant N\leqslant 116$, the 
branching ratio decreases from values close to the SU(3) limit of 
0.7 to the U(5)/O(6) limit of zero. This behavior corresponds to the 
transition from well deformed to $\gamma$-soft nuclei as confirmed 
by the experimental data on Os and W isotopes. At this point, one 
can observe the increase from $N=116$ toward the shell closure 
$N=126$. The increase represents the deviation from the $\gamma$
-soft character, as the corresponding mapped energy surface in Fig.~
\ref{fig:ibm-pes} exhibits notable oblate deformation. The change in 
the branching ratio occurs more slowly than the D1S case 
\cite{IBMwos}. This is consistent with our general finding that the D1M 
energy surfaces for these nuclei show less pronounced quadrupole correlation 
than the D1S ones. As observed in Fig.~\ref{fig:BE2}(b), the branching 
ratios for Pt isotopes remain always much closer to zero, 
which is compatible with their sustained $\gamma$-soft character.

\section{Summary}
\label{sec:summary}

In summary, the method of deriving the Hamiltonian of the interacting
boson model from the constrained HFB
calculations with the Gogny functional D1M has been applied to the
spectroscopic analysis of the neutron-rich Yb, Hf, W, Os, and Pt
isotopes. 
The microscopic energy surface obtained from the
constrained HFB calculation turns out to be a good starting point for
both reproducing and predicting the ground-state shape of the considered nuclei. 
Spectroscopic observables that characterize the 
underlying shape transitions, such as excitation energies, $B$(E2) ratios and
correlation energies, have been calculated.

It has been shown that the Pt isotopes largely differ from the other 
isotopes in the rapidity of the shape transition. For most of the 
considered Pt nuclei the mapped IBM energy surfaces are $\gamma$ 
soft. The transition occurs more rapidly when departing from $Z=76$ 
(Os) through $Z=70$ (Yb). The triaxial deformation helps to 
understand the prolate-to-oblate shape transition that occurs in the 
considered isotopes. The $N=116$ nuclei can be commonly identified 
as the transition points. This is most noticeably seen in the 
overall systematic trend of the band-head of the $\gamma$ band 
$2^{+}_{2}$, as well as in energy and  $B$(E2) ratios. Predicted 
spectra have been presented for the neutron-rich Yb and Hf isotopes, 
where a quite rapid structural evolution is suggested. When compared 
to the results from the standard Gogny-D1S parametrization 
\cite{GognyIBMPt,IBMwos}, the D1M functional seems to be equally valid to 
describe the physics involved.

On the other hand, the present work aims at investigating the 
possible ways of refining the current model and clarifying its 
limitations when applied to the considered mass region. First, as 
discussed in Sec.~\ref{sec:level}, the discrepancy in the level 
structure of the quasi-$\gamma$ band turns out to be a major 
limitation. It is likely that this discrepancy is mainly due to the 
use of the IBM Hamiltonian not reproducing the triaxial energy 
minimum. A specific three-body (cubic) term may improve the 
agreement. Second, the boson effective charges need to be determined 
in a microscopic way and effects beyond the mean 
field, like core polarization, should be taken into account.
It would also be 
meaningful to compare the spectra and the electromagnetic transition 
rates resulting from the present method directly with those obtained 
from  full configuration-mixing and symmetry-conserving 
calculations including triaxial degrees of freedom. This would 
help to quantify the predictive power of the employed model when 
applied to heavy exotic nuclei. Work along these directions is in progress.

\section*{Acknowledgments \label{sec:acknowledge}}
\addcontentsline{toc}{chapter}{Acknowledgments}

This work has been supported in part by grants-in-aid for Scientific
Research (A) 20244022 and No.~217368. 
Author K.N. acknowledges the support by the JSPS. 
The work of authors R.R., L.M.R, and P.S has been supported
by MICINN (Spain) under research grants 
FIS2008--01301, FPA2009-08958, and FIS2009-07277, as well as by 
Consolider-Ingenio 2010 Programs CPAN CSD2007-00042 and MULTIDARK 
CSD2009-00064. Author R.R. thanks Profs. 
J. \"Aysto and R.Julin as well as the  experimental 
teams of the University of Jyv\"askyl\"a (Finland) for 
warm hospitality and encouraging discussions.


\begin{thebibliography}{99}
\bibitem{BE2note}
The experimental $B$(E2) value for $2^{+}_{2}\rightarrow 2^{+}_{1}$ transition in $^{194}$Pt nucleus has
been updated since the one presented in Fig.~4(a) of
Ref.~\cite{GognyIBMPt}. 
Thus, the present work presents more complete compilation of the
experimental $B$(E2) data, taken from Ref.~\cite{BE2}.
\end{thebibliography}
\end{document}